\documentclass[english,aps,preprint,nofootinbib,unsortedaddress,superscriptaddress]{revtex4}
\usepackage[T1]{fontenc}
\usepackage[latin9]{inputenc}
\usepackage{color}
\definecolor{note_fontcolor}{rgb}{0.80078125, 0.80078125, 0.80078125}
\usepackage{verbatim}
\usepackage{amsmath}
\usepackage{amsmath,amssymb,amsfonts,amsthm,cancel,wasysym}
\usepackage{graphicx}
\usepackage{booktabs}

\makeatletter

\newenvironment{lyxgreyedout}
  {\textcolor{note_fontcolor}\bgroup\ignorespaces}
  {\ignorespacesafterend\egroup}

\@ifundefined{textcolor}{}
{%
 \definecolor{BLACK}{gray}{0}
 \definecolor{WHITE}{gray}{1}
 \definecolor{RED}{rgb}{1,0,0}
 \definecolor{GREEN}{rgb}{0,1,0}
 \definecolor{BLUE}{rgb}{0,0,1}
 \definecolor{CYAN}{cmyk}{1,0,0,0}
 \definecolor{MAGENTA}{cmyk}{0,1,0,0}
 \definecolor{YELLOW}{cmyk}{0,0,1,0}
 }

\makeatother

\usepackage{babel}

\begin{document}

\global\long\def\abs#1{\left| #1 \right| }
\global\long\def\half{\frac{1}{2}}
\global\long\def\partder#1#2{\frac{\partial#1}{\partial#2}}
 \global\long\def\comm#1#2{\left[ #1 ,#2 \right] }

\global\long\def\Tr#1{\textrm{Tr}\left\{  #1 \right\}  }

\global\long\def\Imag#1{\mathrm{Im}\left\{  #1 \right\}  }

\global\long\def\Real#1{\mathrm{Re}\left\{  #1 \right\}  }

\global\long\def\db{\!\not\!\! D\,}

\global\long\def\gesim{\,{\raise-3pt\hbox{$\sim$}}\!\!\!\!\!{\raise2pt\hbox{$>$}}\,}
\global\long\def\then{{\quad\Rightarrow\quad}}
\global\long\def\lcal{{\cal L}}
\global\long\def\mcal{{\cal M}}
\global\long\def\bBB{{\mathbbm B}}
\global\long\def\sigbf{\bm{\sigma}}
\global\long\def\gev{\hbox{GeV}}
\global\long\def\tev{\hbox{TeV}}
\global\long\def\VEVof#1{\left\langle #1\right\rangle }
\global\long\def\up#1{^{\left( #1 \right) }}
\global\long\def\inv#1{\frac{1}{#1}}
\global\long\def\su#1{{SU(#1)}}
\global\long\def\ui{U(1)}

\preprint{
   {\vbox {
      \hbox{\bf CAFPE-183/14}
      \hbox{\bf UG-FT-313/14}
      \hbox{\bf CERN-PH-TH-2014-237}
      \hbox{\bf DESY 14-228}
                    }}}

\vspace*{1.5cm}

\title{
Collider limits on leptophilic interactions
}

\author{Francisco del Aguila}

\affiliation{CAFPE and Departamento de F{\'\i}sica Te{\'o}rica y del Cosmos, Universidad
de Granada, E\textendash{}18071 Granada, Spain}

\author{Mikael Chala}

\affiliation{CAFPE and Departamento de F{\'\i}sica Te{\'o}rica y del Cosmos, Universidad
de Granada, E\textendash{}18071 Granada, Spain}

\affiliation{DESY, Notkestrasse 85, 22607 Hamburg, Germany}

\author{Jose Santiago}

\affiliation{CAFPE and Departamento de F{\'\i}sica Te{\'o}rica y del Cosmos, Universidad
de Granada, E\textendash{}18071 Granada, Spain}

\affiliation{CERN, Theory Division, CH1211 Geneva 23, Switzerland}

\author{Yasuhiro Yamamoto}

\affiliation{CAFPE and Departamento de F{\'\i}sica Te{\'o}rica y del Cosmos, Universidad
de Granada, E\textendash{}18071 Granada, Spain}

\begin{abstract}

\vspace*{1.5cm}

Leptophilic interactions can only be observed at the LHC in
four-lepton final states.
If these interactions are mediated by a resonance in the di-leptonic
channel with renormalizable couplings, the mediator must have
spin 1.
We study the LHC reach for such a vector boson allowing for arbitrary
couplings. We find that only couplings to muons can be probed at the LHC  
because lepton flavor violating
couplings  
are constrained by rare processes, couplings to electrons by 
LEP and the LHC 
is not sensitive to final states involving taus in this case. 
The ILC becomes then complementary to the LHC
as it will provide the best limits on 
$Z'$ couplings to tau leptons.
A prominent example is the case of the anomaly-free $Z'$ coupling to the muon 
minus tau lepton number ${\rm L}_\mu - {\rm L}_\tau$. 
If no departure from the Standard Model is observed at the LHC, 
the most stringent bounds on this vector boson are provided from events 
with only three charged leptons plus missing energy. Masses of the
order of 1 TeV can be probed at the
high-luminosity phase of the LHC for $Z^\prime$ couplings of order one.
Generic four-lepton operators parametrizing leptophilic interactions 
can be also constrained using three and four (or two at the ILC)
charged-lepton samples,  
but the corresponding limits are marginal, if meaningful, because the 
resonant behavior appears to be essential for the signal to be significant. 
\end{abstract}

\maketitle

\section{Introduction}
\label{introduction}

The Large Hadron Collider (LHC) is the largest discovery 
machine ever built. 
So far only the last particle within the Standard Model
(SM), the Brout-Englert-Higgs (BEH) boson~\cite{Englert:1964et}, 
has been observed for the first time 
\cite{Aad:2012tfa}. 
Nevertheless, the outstanding performance of the LHC Collaborations 
has allowed to set stringent limits on many different SM extensions, 
especially on strongly interacting particles coupling directly 
to the initial partons. 
For instance, present bounds on dijet resonances are in the 3-5 TeV
region~\cite{Chatrchyan:2013qha}, bounds on quark
compositeness are of the order of $2.5$ TeV~\cite{Khachatryan:2014hpa}
and bounds on new quarks around $800$ GeV~\cite{Aad:2014efa},
respectively 
\footnote{Bounds on gluinos and squarks involve more partons in the 
final state and are more model dependent, being of the order of 1 TeV 
or near this value, repectively \cite{Chatrchyan:2013wxa}.}. 
Similar bounds, however, apply to resonances coupling with 
Electro-Weak (EW) strength to quarks and leptons due to the smaller SM 
backgrounds when the final state is not purely hadronic and includes 
isolated leptons with large transverse momentum. 
For instance, present direct bounds for a new $Z'$ 
contributing to Drell-Yan production, 
$q\bar q \rightarrow Z' \rightarrow l^+ l^-$, 
can be up to 3 TeV for popular SM additions
\cite{Chatrchyan:2012oaa},  
as no departure from the SM prediction has been observed yet. 
Obviously, a hadron machine with an excellent lepton 
reconstruction is the proper place to look for 
New Physics (NP) characterized by resonances coupling sizably to both, 
quarks and leptons, or by the corresponding four-fermion effective 
operators if these new particles are banished to higher energies 
\cite{delAguila:2010mx,delAguila:2011zs,Blas:2013ana} 
(see also \cite{deBlas:2013qqa}). 
If so, little room will be left by the LHC for the discovery  
of new mediators of the reverse process, $l^+ l^- \rightarrow q\bar q$, 
at future lepton colliders, if their Center of Mass Energy (CME) 
is moderate, as in the case of the International Linear Collider
(ILC) \cite{ILC} 
\footnote{The ILC is foreseen as a precision machine and although its CME 
will not allow to produce relatively heavy particles, it will be able
to provide  
indications or indirect constraints on NP well above its production threshold. 
A prime example is the case of an extra $Z'$, for which the diagnostic reach 
at the ILC can be in general larger than at the LHC \cite{DelAguila:1993rw}. 
However, although the discriminating power between different $Z'$ additions 
can be better at the ILC, especially if polarized beams are available and for a 
higher CME, it will be a challenging task 
to establish the existence of a new vector boson coupling to both 
quarks and leptons at this machine if no signal is 
observed at the LHC with a high luminosity. See \cite{Freitas:2013xga} 
for a review and further references.}. 

In this paper we want to address an alternative question: What 
can we learn at large (hadron) colliders about NP which couples only to
leptons?
Or in other words, 
which are the prospects to discover or to exclude leptophilic 
interactions at the LHC and/or ILC?
The departures from the SM predictions in such a case have to be small, 
at least at the LHC, 
because the new processes must involve final leptons to which to attach 
the leptophilic interaction and they are produced with EW strength 
(in the SM) at a hadronic machine. 
In order to enhance the signal significance we may also 
require that it resonates in the di-lepton channel.
This defines the most favorable scenario and we shall discuss it first. 

Such a leptophilic particle must be a boson, for it 
couples to two fermions.  
Moreover, assuming renormalizable coulings, the new particle must be a
vector boson if the SM gauge symmetry and matter content fully describe 
physics below the EW scale, as the LHC data seem to indicate.
Indeed, a scalar multiplet transforming trivially under Lepton Number 
(LN) transformations can only couple to two SM lepton multiplets 
in a renormalizable way as does the SM scalar doublet but 
then, it also couples to the EW gauge bosons, and not only to leptons, 
as we assume. 
If we obviate this possibility by requiring the scalar to be a neutral singlet, 
we can only couple it to lepton pairs through non-renormalizable terms, 
which are in general effectively suppressed by small lepton masses 
because the corresponding lepton product must involve two multiplets with 
the same (wrong) chirality or an extra BEH boson insertion. 
Finally, if the new scalar multiplet has non-zero LN (in fact, equal to 2), 
it must have a doubly-charged component, coupling again to the 
SM gauge bosons, too 
(see Ref. \cite{delAguila:2013yaa} 
for a detailed discussion of the lowest 
order couplings and production mechanisms of these extra scalars).
In summary, the only leptophilic particle with renormalizable 
couplings to SM lepton pairs is a new neutral vector boson $Z'$ 
\footnote{Obviously, such a vector boson can mix with the Z boson 
\cite{Holdom:1985ag} 
and hence, also contribute to Drell-Yan production. As a matter of 
fact, this mixing is generated by quantum corrections in models 
with generic couplings if the mixing term is not already present after
integrating  
out the heavy modes of a more fundamental theory at higher energies 
\cite{delAguila:1988jz}.
Such a mixing is in general small, as already experimentally required 
by present LHC bounds on new vector bosons contributing to Drell-Yan 
production \cite{Chatrchyan:2012oaa} 
(see also~\cite{Andreev:2014fwa},
and references there in). These limits then make negligible its 
contribution to four-fermion $Z'$ final states and hence, we can 
neglect this mixing throughout the paper when studying the leading 
contribution to four-lepton production mediated by a 
leptophilic vector boson. (See Ref. \cite{Corfu2014} for a more detailed 
discussion.)} 
with the following interaction Lagrangian:  
\begin{equation}
{\cal L}_{Z'} = - (g^{\prime ij}_{\rm L} {\overline L_{{\rm L} i}} \gamma^\mu L_{{\rm L} j}  + 
g^{\prime ij}_{\rm R} {\overline l_{{\rm R} i}} \gamma^\mu l_{{\rm R} j}) Z'_\mu \ ,
\label{zprimecouplings}
\end{equation}
where 
$g^{\prime ij}_{{\rm L, R}}$ are arbitrary dimensionless couplings to 
the SM Left-Handed (LH) and Right-Handed (RH) lepton multiplets, 
$L_{{\rm L} i} = \left( \begin{array}{c}  \nu_{{\rm L} i} \\ l_{{\rm L} i} \end{array} \right)$
and $l_{{\rm R} i}$, respectively, with $i = e, \mu, \tau$ labeling
the charged-lepton  
family 
\footnote{Thus, although arbitrary flavor and chiral interactions are
  allowed, the EW gauge symmetry remains unbroken for the new
  couplings of the LH charged leptons and their neutrino counterparts
  are equal. However, the SM Yukawa couplings do not preserve such a
  hypothetical gauge symmetry. As a matter of fact, the extra U(1)
  symmetry coupling to muon minus tau lepton number discussed below
  allows for Yukawa couplings which do preserve the chiral but not the
  flavor symmetry \cite{Foot:1994vd}. But in general, one can build
  realistic models with extra dimensions or with strong EW symmetry
  breaking (in which the degree of compositeness of LH and RH leptons
  does not have to be equal) which predict massive vector bosons with
  flavor and chiral dependent interactions. (See
  \cite{delAguila:2010vg,delAguila:2010es} for an example in composite
  Higgs models in which the RH tau lepton couples differently than the
  LH one to the extra heavy vector bosons present in the spectrum.) In
  this case, once they acquire a mass, Yukawa couplings are generated
  at higher orders in perturbation theory.}
Higher-spin particles do not have renormalizable couplings to lepton pairs either.

\begin{figure}
\begin{centering}
\hspace{1cm}\includegraphics[width=0.65\columnwidth]{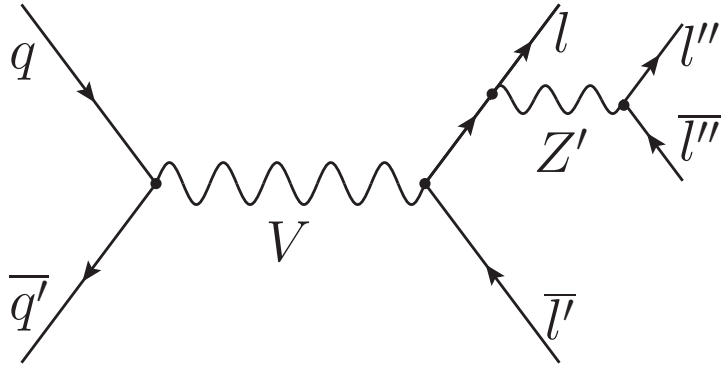}
\end{centering}
\caption{Leading production diagram of a leptophilic $Z'$ at the LHC.
\label{zprimeproduction}
}
\end{figure}

In the following we shall discuss the phenomenology of such a 
vector boson, leaving to the end the comments on the case where it is
beyond the LHC reach and its effects are parametrized 
by the corresponding four-lepton operators at low energy. 
There are different NP sources for four-lepton signals at large colliders, 
as, for instance, new doubly-charged scalars introduced above or heavy neutrinos 
decaying into three leptons, 
with also different production mechanisms. 
Thus, whereas in the former case the dominant contribution results 
from EW pair production \cite{delAguila:2013yaa}, in the latter it is through fermion 
mixing \cite{Han:2006ip}, but also mediated by gauge boson exchange. 
In our case the leading contribution for the production of a leptophilic 
vector boson at the LHC is illustrated in Fig. \ref{zprimeproduction}, 
with the $Z'$ emitted from one of the final leptons in Drell-Yan production. 
The exchanged EW gauge boson can be also a $W$ if 
$g^{\prime}_{{\rm L}}$ is sizable. As a matter of fact, in general the 
corresponding process with only three charged leptons in the final state 
plus missing energy provides the most stringent limit on a leptophilic $Z'$.

Large collider bounds on the couplings of such a vector boson 
are in general effectively less restrictive than indirect limits derived 
from precision experiments and hence, we must review the latter before 
going on. 
All processes constraining a leptophilic vector boson involve at 
least four leptons and two vertices, internal or not. Hence, a small 
contribution can be the result of a small coupling squared or of one 
smaller coupling multiplying a larger one. This makes the Large Electron-Positron collider (LEP) 
constraints very stringent, when they apply. 
The $Z'$ couplings to electrons are mainly 
constrained by the very precise $e^+e^- \rightarrow e^+e^-$ LEP data,
that very closely follows the SM prediction~\cite{Blas:2013ana} 
\footnote{These 95 \% Confidence Level (C.L.) bounds are obtained assuming universality 
but there is no a large flavor dependence \cite{Schael:2013ita} and hence, 
they can be considered a proper estimate of the present upper limits.}, 
leaving no room for further improvement at the LHC: 
\begin{equation}
\frac{g^{\prime ee}_{\rm L,R}}{M_{Z'}} < 0.12, 0.16\ {\rm TeV}^{-1}\;  .
\label{zprimeelectroncouplings}
\end{equation}
Similar limits can be derived for $g^{\prime e\mu, e\tau}_{\rm L,R} / M_{Z'}$ 
from $e^+e^- \rightarrow \mu^+\mu^-, \tau^+\tau^-$, respectively. 
As a matter of fact, the actual limits are slightly more stringent 
in this case but these bounds are already restrictive enough to make the 
corresponding vector boson production unobservable at the LHC, 
as it will be apparent from the analyses of a $Z'$ mainly coupling to muons 
in the next section.  
On the other hand, bounds on Lepton Flavor Violating (LFV) muon decays
like, for instance,  
$\Gamma (\mu \rightarrow 3e) /  \Gamma (\mu \rightarrow {\rm all}) < 10^{-12}$ 
at the 90 \% C.L. \cite{Bellgardt:1987du} 
can be fulfilled assuming the diagonal or the off-diagonal 
coupling to be negligible, $g^{\prime ee}_{\rm L,R} / M_{Z'}$ 
or $g^{\prime e\mu}_{\rm L,R} / M_{Z'}$, in this example. 
In any case, however, the previous bounds are in practice restrictive enough 
to neglect them in the study of the LHC and ILC reach for a leptophilic $Z'$. 
Finally, the limit on 
$\Gamma (\tau \rightarrow 3\mu) /  \Gamma (\tau \rightarrow {\rm
all}) < 2.1 \times 10^{-8}$ at the 90 \% C.L. \cite{Hayasaka:2010np} implies 
$g^{\prime \mu\mu}_{\rm L,R} g^{\prime \mu\tau}_{\rm L,R} / M^2_{Z'} <
10^{-2} ~{\rm TeV}^{-2}$,  
which can be only satisfied requiring a small enough 
$g^{\prime \mu\tau}_{\rm L,R} / M_{Z'}$ because the vector boson must be 
reconstructed through its decays to muon pairs to be observable at the LHC 
and then, the diagonal muon couplings can not be too small.
Obviously, a proper discussion of all these limits would require to 
disentangle all possible products of $Z'$ couplings contributing 
to the different processes, also using their angular distributions, 
but the conclusion would be the same.  
Hence, only the diagonal $Z'$ couplings to muons and taus 
can be eventually further constrained at the LHC, assuming in either case small 
enough couplings to electrons in order to satisfy the LEP (and eventually ILC) limits on  
$e^+e^- \rightarrow e^+e^-, \mu^+\mu^-, \tau^+\tau^-$. 
We will denote these four coupling constants with one upper index at most, 
$g^{\prime \mu, \tau}_{{\rm L,R}}$, from now on. 

As already emphasized, in order to observe a leptophilic $Z'$ at the LHC one 
has to sample four-lepton events,  
$q\bar q \rightarrow l \bar l l\bar l$,  as this vector boson must be 
emitted from a lepton, and decay afterwards into a lepton pair 
(see Fig. \ref{zprimeproduction})
\footnote{Radiation of EW gauge bosons by one of the final leptons in Drell-Yan 
production was proposed some time ago to devise new observables which 
could help to further characterize heavy $Z'$s contributing to this process 
\cite{Cvetic:1992qv} (see also \cite{delAguila:1992ap}). 
In our case, however, it is the $Z'$ itself which is emitted from the final lepton in 
Drell-Yan production, whereas the EW gauge bosons are exchanged in the $s-$channel. 
In both processes the radiated vector boson has the tendency to align along the emitting 
lepton and hence, opposing to the other one with larger momentum, in general.}. 
Depending on the value of the coupling constants $g^{\prime \mu, \tau}_{{\rm L,R}}$, 
$l$ can be a muon or a tau or one of their neutral counterparts. 
However, only the samples with 4$\mu$ or 3$\mu$ plus missing energy 
with one $\mu^+ \mu^-$ pair reconstructing the $Z'$ mass are sensitive to such a 
vector boson at the LHC, as we shall show. 
Otherwise, the small branching ratios or the large irreducible 
backgrounds from gauge boson pair production make the 
signal unobservable. 

We shall discuss in the following how to proceed in order to characterize 
the new vector boson at the LHC and the ILC. 
The general case is fixed, as we have stressed by the $Z'$ 
mass, $M_{Z'}$, its four couplings to muons and taus, 
$g^{\prime \mu, \tau}_{{\rm L,R}}$, and its total width, $\Gamma_{Z'}$. 
Obviously, the latter is the sum of its partial decay rates into muons 
and taus (and their associated neutrinos), 
\begin{equation}
\Gamma_{Z'}^{\mu + \tau} = 
\frac{2 g^{\prime \mu\, 2}_{{\rm L}} + g^{\prime \mu\, 2}_{{\rm R}} 
+ 2 g^{\prime \tau \,2}_{{\rm L}} + g^{\prime \tau \,2}_{{\rm R}}}{24 \pi} 
M_{Z'}
\; ,
\label{zprimepartialwidth}
\end{equation}
and into any other channel which may be open. 
Thus in general, $\Gamma_{Z'} \geq \Gamma_{Z'}^{\mu + \tau}$ is an 
extra free parameter to be determined experimentally, typically 
adjusting the corresponding Breit-Wigner distribution. 
As the LHC is only sensitive to the muon signals, as already pointed out 
and proven in next section, we will only vary 
$M_{Z'}$ and $g^{\prime \mu}_{{\rm L,R}}$ in the corresponding 
analyses, assuming that $\Gamma_{Z'} = \Gamma_{Z'}^{\mu}$. 
In this case the three free parameters can be determined comparing 
the 4$\mu$ and 3$\mu$ samples. 
In order to estimate the effect of extra decay channels and hence 
of a larger $\Gamma_{Z'}$, 
let us discuss, for illustration, the case of 
the only anomaly-free $Z'$ addition in this class of models 
\cite{Foot:1994vd}, the combination of muon minus tau LN, 
$Z'_{\mu - \tau}$, with charges $g^{\prime \mu, \tau}_{{\rm L,R}}/g'_{\rm L}$ 
given by 
\begin{equation}
\begin{tabular}{l|cccc}
Multiplet \quad & 
\quad $L_{{\rm L} \mu} = \left( \begin{array}{c}  \nu_{{\rm L} \mu} \\ \mu_{\rm L} \end{array} \right)$ \quad & 
\quad $\mu_{\rm R}$ \quad & 
\quad $L_{{\rm L} \tau} = \left( \begin{array}{c}  \nu_{{\rm L} \tau} \\ \tau_{\rm L} \end{array} \right)$ \quad & 
\quad $\tau_{\rm R}$ \quad  \\
\hline
Charge & 
$1$ & 
$1$ & 
$- 1$ & 
$- 1$
\end{tabular}
\nonumber
\end{equation}
In the narrow width approximation \cite{narrowwidth}, 
which is a good prescription at the LHC (ILC) up to tens of per cent in the less 
favorable case of very large vector couplings, this model gives the same predictions 
for 3$\mu$ plus missing energy and 4$\mu$ 
production as the model with vanishing couplings to taus 
but with vector couplings to muons a factor of $\sqrt 2$ smaller. 
This reflects the fact that in this approximation the production cross-sections 
scale as $\Gamma_{Z'}^{-1}$ and $\Gamma_{Z'}^{\mu + \tau} = 2 \Gamma_{Z'}^{\mu}$, 
and this global factor of 1/2 can be absorbed in the redefinition of the strength of the 
$Z'$ emission from one of the final leptons in Drell-Yan production (Fig. \ref{zprimeproduction}).
In general, the sign of the charges does not play any role in this case because there is no interference 
between different final states, and no sign determination is possible either through 
the production processes studied here 
\footnote{This is so, however, for the case of a resonant mediator of the leptophilic interaction and 
when the SM background is relatively small but not, for example, for SM extensions involving four-lepton 
operators, as we shall discuss in the following. We will find, for instance, that their interference with 
the SM background makes the sign of their coefficients physically meaningful.}. 

The couplings for the different cases (set of $Z'$ couplings) have been 
included in a Universal Feynman rules Output 
(\texttt{UFO}) model \cite{Degrande:2011ua} 
by means of \texttt{FEYNRULES} \cite{Christensen:2008py}. 
It can be downloaded from 
{\textsf{http://cafpe.ugr.es/index.php/pages/other/software}}. 
A set of input parameters (\texttt{Param Cards}) for 
\texttt{MADGRAPH 5} \cite{Alwall:2011uj} can be also found there for
all $Z'$ masses  
considered in the simulations along the text. 

A few comments are in order. The LHC limits for $Z'_{\mu - \tau}$ have been 
recently discussed in the literature
\cite{Heeck:2011wj,Harigaya:2013twa}  
(see also
\cite{Bell:2014tta} for the analysis of a related model with
leptophilic dark matter) but concentrating 
on the 4$\mu$ channel, which provides the best bounds only for
$g^{\prime \mu}_{{\rm R}}$  
somewhat larger than $g^{\prime \mu}_{{\rm L}}$. Otherwise, the 3$\mu$ plus 
missing energy channel provides the most stringent limits, as suggested in 
\cite{Ma:2001md} and we confirm by a detailed analysis in the
following. 
On the other hand, independent bounds from neutrino trident production 
also impose stringent constraints on $Z'_{\mu - \tau}$ and on any
other $Z'$ model  
with a non-vanishing coupling $g^{\prime \mu}_{{\rm L}}$
\cite{Altmannshofer:2014cfa}.  
Nevertheless, they still leave room for further improvement at the LHC
and the ILC. In particular, no constraint on $g^{\prime \mu}_{{\rm R}}$ is 
set by $\mu^+ \mu^-$ production from the scattering of the muon neutrino 
off the Coulomb field of a nucleus. 
The constraints from $g-2$ are weaker but also apply to the RH muon coupling~
\cite{Harigaya:2013twa,Altmannshofer:2014cfa}. 
Finally, we focus our attention on $Z'$ masses above the $Z$ mass, 
as we want to investigate the ultimate reach at large colliders. 
Masses below $M_Z$ have been considered in~
\cite{Harigaya:2013twa,Altmannshofer:2014cfa,Bell:2014tta}.

In next section we study the LHC reach for a leptophilic $Z'$. 
We derive discovery and exclusion limits as a function of its mass, $M_{Z'}$, 
and its LH coupling to muons, $g'_{\rm L} \equiv g^{\prime \mu}_{\rm L}$, for 
different values of the corresponding RH coupling, $\xi \equiv g^{\prime \mu}_{\rm R}/g^{\prime \mu}_{\rm L}$. 
Such a vector boson could be eventually excluded at the LHC 
\footnote{The limits derived using Run I data are much weaker than 
the indirect bounds from neutrino trident production, $\nu N \rightarrow \nu \mu^+ \mu^- N$.}
for $Z'$ masses up to $\sim 1$ TeV 
and $g'_{\rm L}$ and $\xi$ of order one in the high-luminosity phase with an 
integrated luminosity ${\cal L}_{\rm int} = 3\ {\rm ab}^{-1}$ and a CME 
$\sqrt s = 14$ TeV. 
If a leptophilic $Z'$ below this mass is discovered, 
its couplings can be fully determined up to a global normalization 
proportional to the inverse of the square root of the total $Z'$ width, 
$\Gamma_{Z'}^{-1/2}$, which has to be measured independently 
from the invariant $\mu^+ \mu^-$ mass distribution reconstructing 
the $Z'$ mass.  
The codes used and the cuts applied are described in the corresponding 
section.
The analyses to study the corresponding ILC reach are described
in Section \ref{ILC}. 
In this case there is no $W$ exchange contribution but the 3$\mu$ plus missing energy 
sample is traded by the 2$\mu$ plus missing energy one, 
obtaining at the end bounds comparable to those derived at the LHC but only 
for low enough $Z'$ masses (smaller than the ILC CME $\sqrt s = 500$ GeV) 
and for an integrated luminosity ${\cal L}_{\rm int} = 500\ {\rm fb}^{-1}$. 
However, at the ILC not only the $Z'$ total width is expected to be 
measurable with a better precision but in contrast with the LHC, 
the $Z'$ couplings to taus can be constrained analyzing   
the 2$\mu$ plus missing energy and 2$\mu$2$\tau$ final states, too. 
Finally, in Section \ref{4loperators} we comment on the limit 
of very large $M_{Z'}$ and arbitrary effective four-lepton 
interactions. Although bounds on their coefficients can be 
also derived from the corresponding production cross-sections, 
they are too weak to allow for a resonance interpretation. 
Section \ref{conclusions} is devoted to conclusions and final remarks.

\section{LHC reach for a leptophilic $Z'$}
\label{4m} 

As argued in the Introduction, current experimental constraints 
set stringent limits on the coupling of a leptophilic $Z'$ to electrons or to a 
pair of leptons of different flavor. 
Thus the LHC can be only 
sensitive, a priori, to its diagonal couplings to muons and taus. 
As a matter of fact, only the couplings to muons can be determined at the 
LHC because the SM backgrounds are too large for taus 
to make any conclusive claim. 
We present in this section a detailed simulation analysis to 
estimate the LHC potential for discovering or excluding a new 
heavy neutral vector boson of mass $M_{Z'}$ coupling to the 
LH muon doublet with strength $g'_{\rm L}$ and to the RH muon 
singlet with strength $\xi g'_{\rm L}$. Its total width reads
\begin{equation}
\Gamma_{Z'} = \frac{g^{\prime 2}_{\rm L} (2 + \xi^2) + W^{i\neq \mu}}{24 \pi} M_{Z'} \ ,  
\label{totalwidth}
\end{equation}
where $W^{i\neq \mu}$ takes into account all decay channels different than 
$\mu^+ \mu^-$ and $\bar \nu_\mu \nu_\mu$.

The main production mechanism is the one depicted in
Fig. \ref{zprimeproduction}  
with the $Z'$ radiated by one of the final muons (neutrinos) in 
Drell-Yan production. 
The production cross-section for such a (leptophilic) $Z'$ is
relatively small and suppressed  
for heavy vector boson masses $M_{Z'}$. Depending on the initial
parton state the  
exchanged EW gauge boson can be neutral ($\gamma, Z$) and the final mode 
4$\mu$ or 2$\mu$2$\nu_\mu$, or charged ($W$) and the final mode
3$\mu$1$\nu_\mu$.  
Only events with at least two muons are considered because we assume that 
$Z' \rightarrow \mu^+ \mu^-$ in order to allow for the
vector boson reconstruction, 
which is compulsory to enhance the signal to background ratio to an
observable level. In practice, we only consider the two most 
promising channels: $3\mu$ plus missing energy and $4\mu$. 

We have implemented the model in a \texttt{UFO} format~\cite{Degrande:2011ua}
using \texttt{FEYNRULES}~\cite{Christensen:2008py}. Parton-level
events have been 
generated with \texttt{MADGRAPH 5}~\cite{Alwall:2011uj} and
showered/hadronized with \texttt{PYTHIA 6}
~\cite{Sjostrand:2006za}. Detector effects have been simulated
with \texttt{DELPHES 3}~\cite{deFavereau:2013fsa} and the experimental
analyses performed by means of \texttt{MADANALYSIS
  5}~\cite{Conte:2012fm}. Jets are reconstructed with the  
anti-$k_t$ algorithm with $R=0.5$, as implemented in
\texttt{FASTJET}~\cite{Cacciari:2011ma}.   

The applied cuts are collected in Tables~\ref{3mu_cuts:table}
and~\ref{4mu_cuts:table} for the $3\mu$ plus missing energy and $4\mu$
final states, respectively.
The $3\mu$ plus missing energy analysis closely follows the one
in~\cite{atlas_conf_2013_035}, which we found to be the most
constraining among the current experimental searches. Still, current
constraints are much weaker than the indirect ones from the neutrino
trident process~\cite{Altmannshofer:2014cfa} and hence, we shall not report 
detailed results at $\sqrt{s}=8$ TeV.
As an example we have run the analysis in~\cite{atlas_conf_2013_035}
for a $Z^\prime_{\mu - \tau}$ with mass of 140 GeV and coupling
$g^\prime_{\rm L}=0.3$ (at the verge of the trident exclusion
bound). This analysis is similar to the one presented in Table
\ref{3mu_cuts:table} except that it does not include the last (Mass
window) cut and the numerical values of the constraints are somewhat
less stringent. The current bound on the observed number of events reported in
the region named ``SRnoZc'' in~\cite{atlas_conf_2013_035} 
(the most sensitive to our signal) is 6.8
whereas we obtain that only 1.5 of our signal events pass those cuts. 
A simple
modification of the experimental analysis requiring only
muons in the final state (as opposed to muons and electrons) and 
including the last (Mass window) cut would reduce the background
to an almost negligible level but also our signal which would be too
small to provide a significant limit. The situation is even worse in the
4$\mu$ channel in which the cross sections are even smaller.
Nevertheless, we have used the $3\mu$ plus missing energy
analysis to validate our generator implementation, finding good agreement.
In all simulations we have only considered the irreducible
backgrounds, which are by far the most relevant ones, 
unless otherwise stated, 
renormalizing our results with a global factor to account for 
non-irreducible ones as well as for higher-order effects. In
order to be conservative, this factor is applied to the 
estimation of the background but not to the signal.
Finally, an important observation is that it would be extremely
useful that the experimental collaborations present their
results separately for electrons and muons as only the latter
contribute to our signal.
\begin{table}[t]
\begin{center}
\begin{tabular}{lcl}\toprule
Basic cuts & \hspace{1cm} & $p_{\rm T}^\ell > 50$ GeV, 
$|\eta_{\ell}|< 2.4 $, $\Delta R(j\ell) > 0.4$, $p_{\rm T}^j > 20$ GeV, $|\eta_j| < 2.5$ \\\midrule
b veto & & No b jets \\
Number of muons & & $N_\mu = 3$ (net charge = $\pm 1$)\\
Low mass resonance veto & & $m_{\mu^+ \mu^-} > 12$ GeV \\
$Z$ veto & & $|m_{\mu^+\mu^-} - M_Z| > 10$ GeV\\
Missing $E_{\rm T}$ & & $\cancel{E}_{\rm T} > 100$ GeV\\
Transverse mass & & $m_{\rm T} > 110$ GeV\\
Mass window & & $|m_{\mu^+\mu^-} -M_{Z'}| < 0.1M_{Z'}$\\
\bottomrule
\end{tabular}
\caption{Cuts for $3\mu$ plus missing energy events. The transverse 
mass is computed with the transverse missing energy and the lepton 
not belonging to the pair which better reconstructs the $Z'$ boson mass. 
Only isolated muons are considered.
\label{3mu_cuts:table}}
\end{center}
\end{table}
\begin{table}[t]
\begin{center}
\begin{tabular}{lcl}\toprule
Basic cuts & \hspace{1cm} & $p_{\rm T}^\ell > 30$ GeV, $\eta_\ell < 2.4$, $\Delta R(j\ell) > 0.4$\\\midrule
Number of muons & & $N_\mu = 4$ (zero net charge)\\
muon spectrum & & $p_{\rm T}^{\mu_{1,2,3}} > 100,~80,~60$ GeV\\
$Z$ veto & & $|m_{\mu^+\mu^-} - M_Z| > 10$ GeV\\
Mass window & & $|m_{\mu^+\mu^-} -M_{Z'}| < 0.1M_{Z'}$\\\bottomrule
\end{tabular}
\caption{Cuts for $4\mu$ events. Only isolated muons are
  considered. \label{4mu_cuts:table}} 
\end{center}
\end{table}

We have generated and analyzed signal events for different masses and
couplings, as well as irreducible SM backgrounds, 
always aiming at generating samples with an integrated luminosity about
five times larger than the target one of 3 ab$^{-1}$ in order to reduce the
Monte Carlo uncertainties. Once the events have been analyzed, we use
the $CL_s$~\cite{Agashe:2014kda} 
method to obtain $95\%$ C.L. bounds on the corresponding
$Z'$ signal. To assess the discovery potential we use
$\mathcal{S}=5$ with 
\begin{equation}
\mathcal{S}(s,b)=\sqrt{2 \left( (s+b)\log
  \left(1+\frac{s}{b}\right)-s\right)} \ ,
\end{equation}
which gives accurate results for the Monte Carlo statistics
used~\cite{Cowan:2010js}.  In
the above equation $s$ and $b$ stand for the number of signal and
background events, respectively.

\begin{figure}
\begin{centering}
\includegraphics[width=0.49\columnwidth]{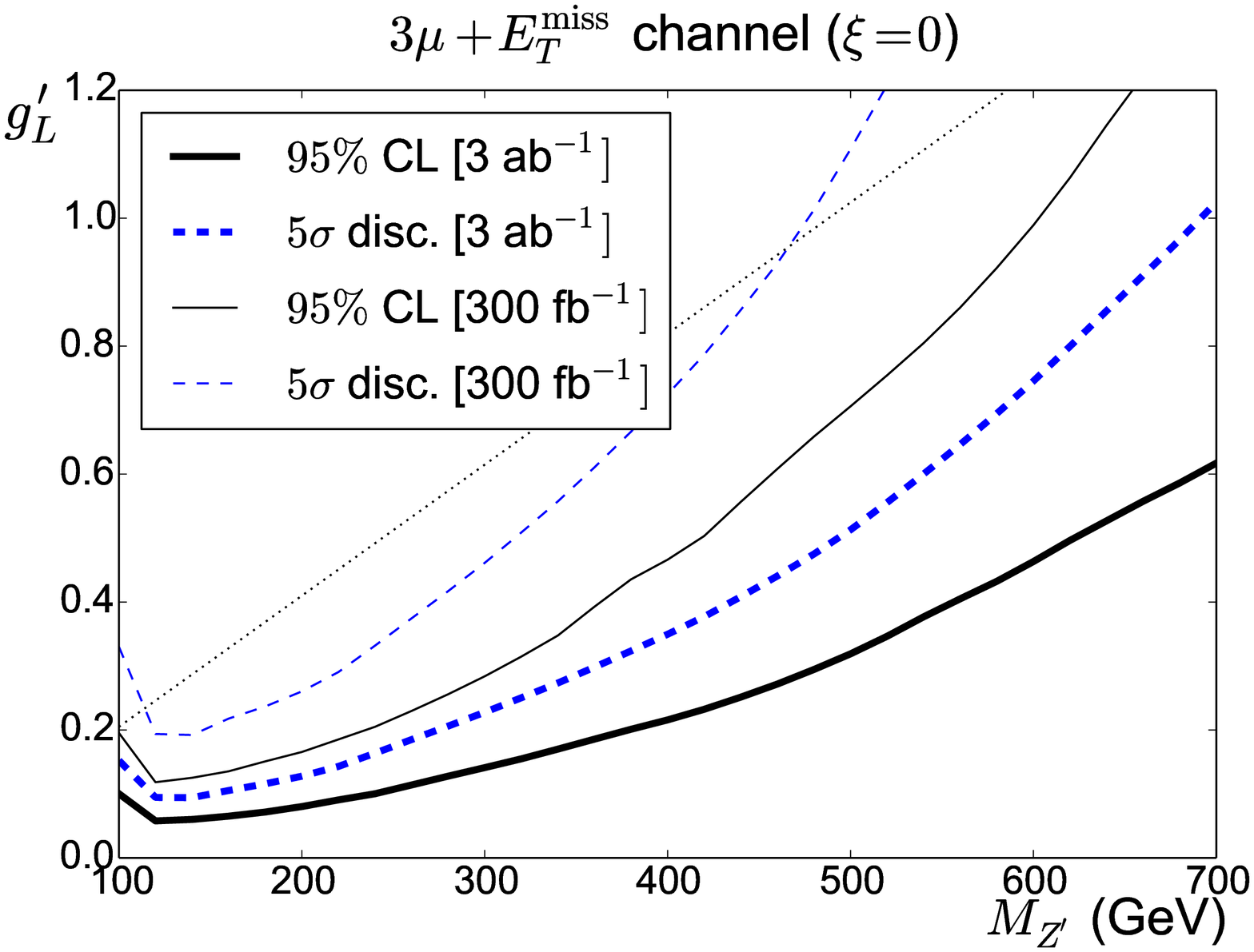}
\includegraphics[width=0.49\columnwidth]{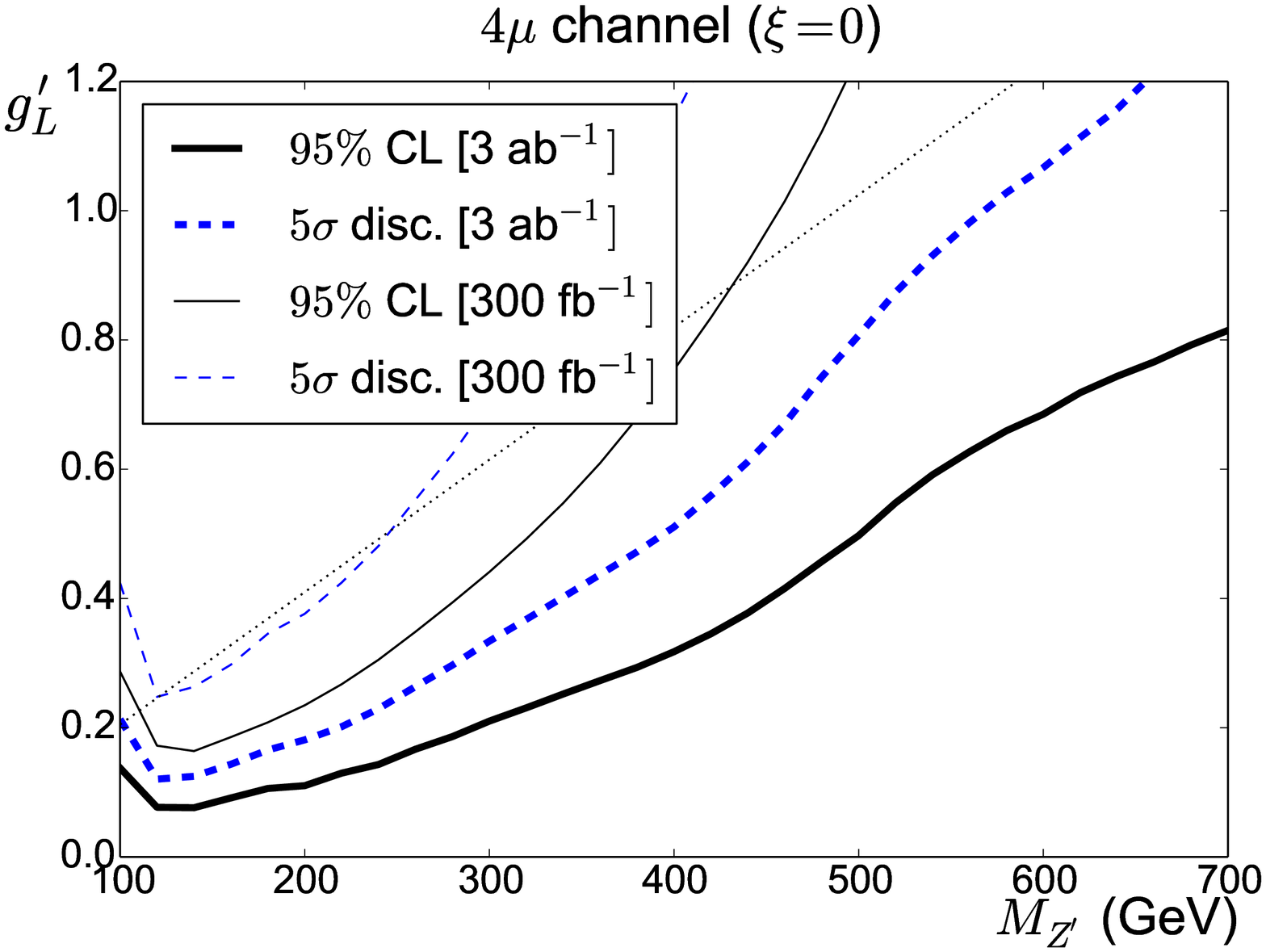}
\par\end{centering}
\caption{Discovery and exclusion limits for a leptophilic $Z'$ only
  coupling to  
LH muons as a function of its mass at the LHC. We also draw the bounds 
from neutrino trident production (straight line), 
for comparison (see the text for details).
\label{Mgxidiscovery}
}
\end{figure}

In Fig. \ref{Mgxidiscovery} we draw the LHC 5$\sigma$ discovery
(dashed curve) and the 95 \% C.L. exclusion (solid curve) limits 
for a $Z'$ only coupling to LH muons as 
a function of its mass for 3$\mu$ plus missing energy (left) and 4$\mu$ (right) events. 
Thick (thin) curves represent the results for an integrated luminosity
of 3 ab$^{-1}$ (300 fb$^{-1}$). 
We also show the bound from the neutrino trident production (dotted line) 
\cite{Altmannshofer:2014cfa} 
taking into account CHARM-II and CCFR data, for comparison. This bound
compares with 
arbitrary $\xi$ curves because it does not depend on this
parameter. Other bounds, like the one from $g-2$ are much weaker,
although they involve both couplings and become more stringent for larger
$\xi$.
As we can see in the Figures, current constraints leave some room
(mainly in the $3\mu$ plus missing energy channel) for
discovery even with the limited integrated luminosity of 300
fb$^{-1}$. Masses up to $\sim$ 1 TeV can
be probed for $g'_{\rm L}$ of order one 
in both channels at the high-luminosity phase of the LHC.

\begin{figure}
\begin{centering}
\includegraphics[width=0.49\columnwidth]{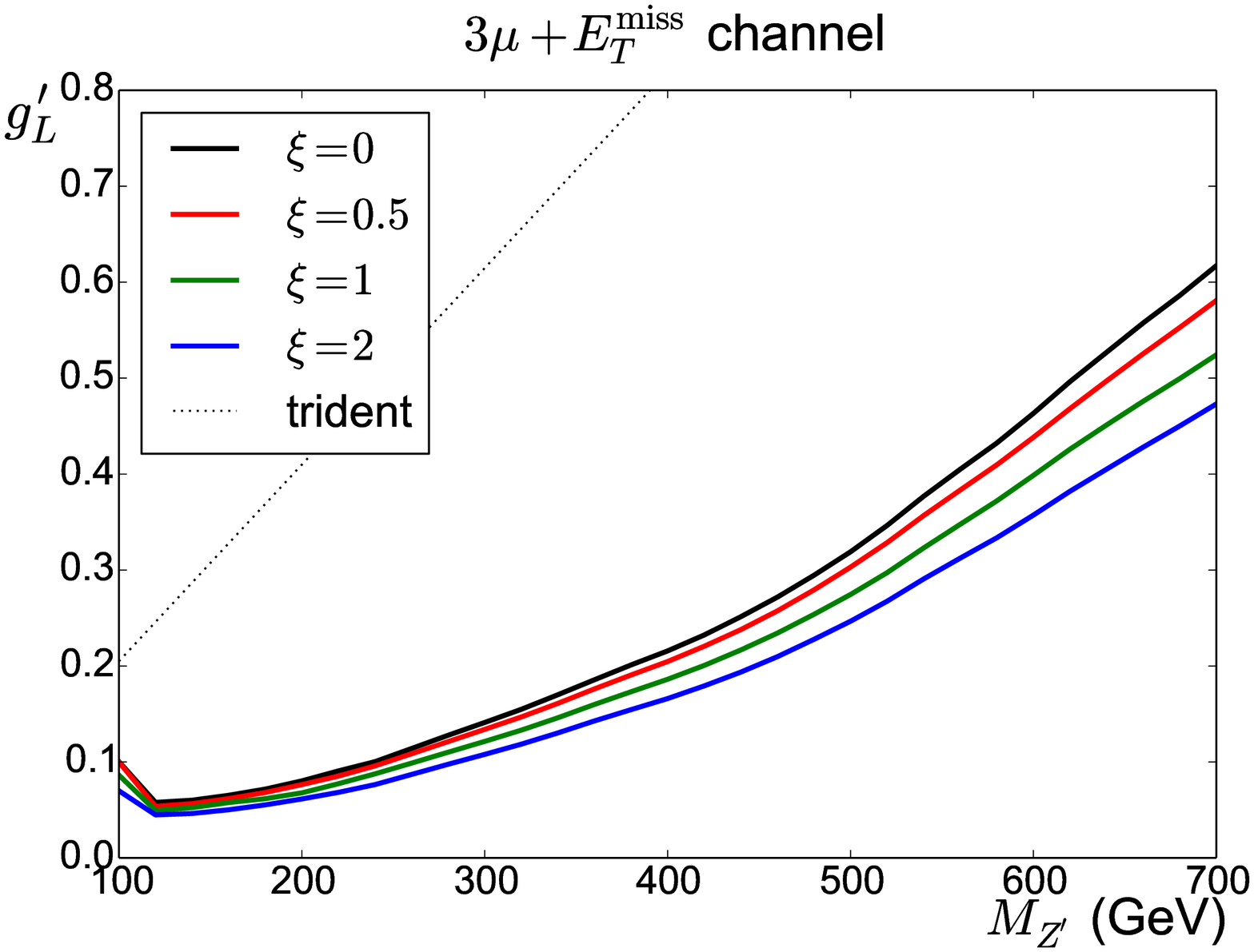}
\includegraphics[width=0.49\columnwidth]{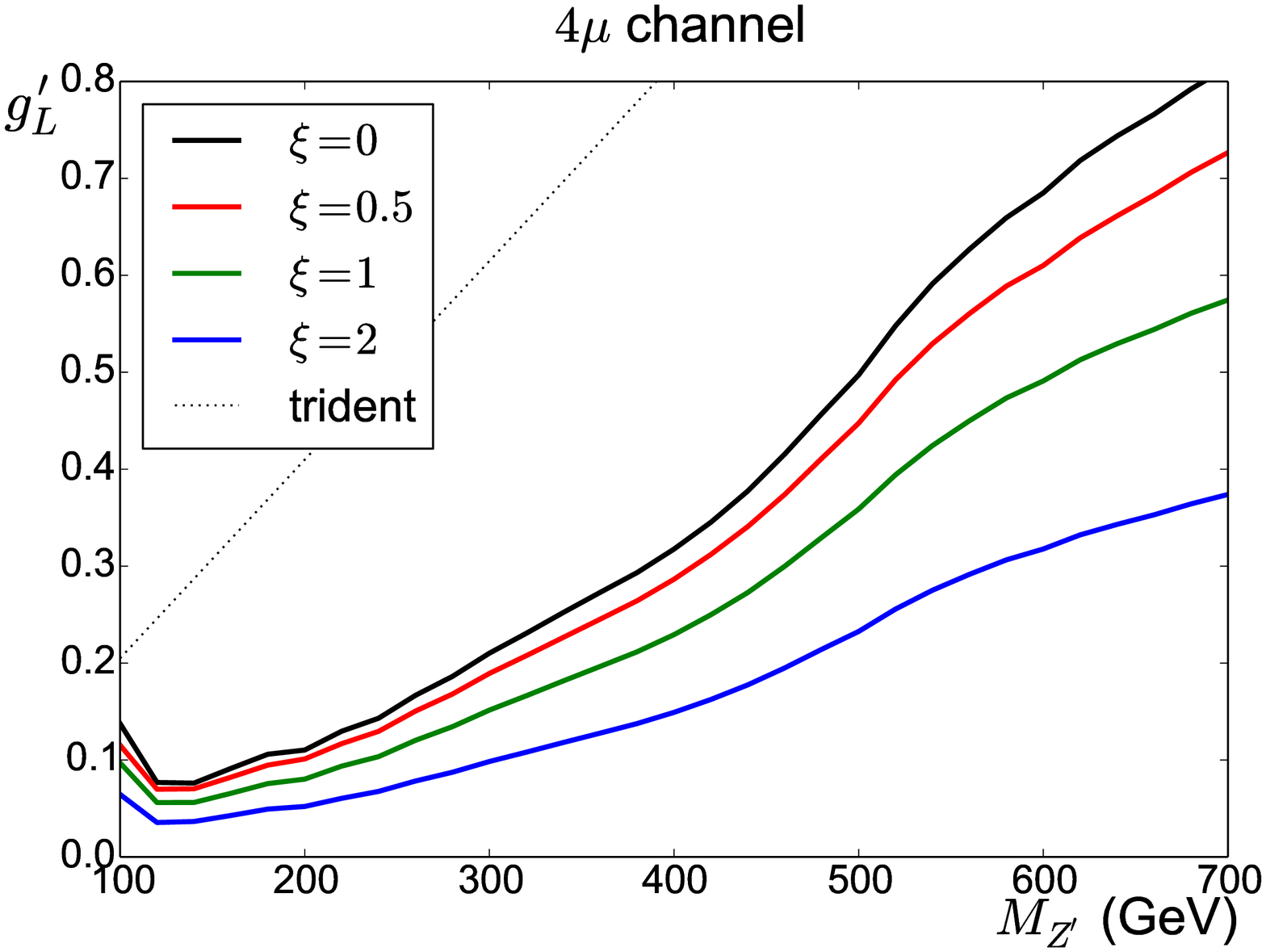}
\par\end{centering}
\caption{
95 \% C.L. exclusion limits for a leptophilic $Z'$ as a function of 
its mass and its coupling to LH muons for different values of its coupling 
to their RH counterparts at the LHC in the $3\mu$ plus missing energy 
(left) and $4\mu$
(right) channels. The different lines correspond to $\xi=0,~0.5,~1$
and $2$, from top to bottom, respectively. The bound from neutrino
trident production (straight line) is also shown for comparison.
\label{Mgxi}
}
\end{figure}

Obviously, the LHC limits also depend on the RH coupling to muons.  
In Fig. \ref{Mgxi} we show the corresponding exclusion limits implied 
by the non-observation of a departure from the SM prediction in the
3$\mu$ plus missing energy (left panel) and $4\mu$ (right panel) channels, as a function
of the vector boson mass $M_{Z'}$ and the $Z'$ coupling to  
LH muons $g'_{\rm L}$ for different values of the RH coupling $\xi g'_{\rm L}$. 
These limits improve with the value of the coupling to RH muons $\xi
g'_{\rm L}$ in  
both cases; but much faster for the neutral final state because the
two exchanged  
EW gauge bosons, $\gamma$ and $Z$, also couple to RH muons, 
which is not the case for the charged final state which requires the
exchange of a $W$ gauge boson which only couples to LH fermions. 
(In this case the contribution proportional to $g'_{\rm R}$ is suppressed by the
ratio of the muon over the $Z'$ mass, being therefore negligible.) 
We assume in all cases that there are no other $Z'$ decay channels open.
The 2$\mu$2$\nu_\mu$ sample is dominated by the irreducible 
background. 

\begin{figure}
\begin{centering}
\includegraphics[width=0.65\columnwidth]{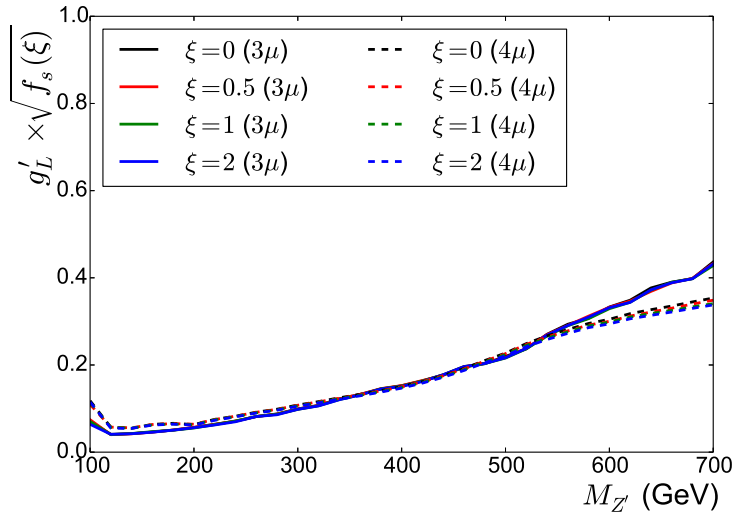}
\par\end{centering}
\caption{95 \% C.L. exclusion limits for a leptophilic $Z'$ as a function of 
its mass and its coupling to LH muons for different values of its coupling 
to their RH counterparts at the LHC, corrected for the new vector
boson coupling dependence  
in Eqs. (\ref{LHCcrosssections}):
$f_{3\mu}(\xi)=\frac{1+\xi^2}{2+\xi^2}$
and $f_{4\mu}(\xi)=(2.5+\xi^2)\frac{1+\xi^2}{2+\xi^2}$.
\label{Limitdependence}
}
\end{figure}

As it is apparent from Figs. \ref{Mgxidiscovery} and \ref{Mgxi}, the limits for the charged final state 
are more stringent than for the neutral one, except for somewhat larger $Z'$ coupling to RH 
than to LH muons and hence, for relatively larger 4$\mu$ cross-sections. 
It is also evident that the dependence on $\xi$ is larger in the 4$\mu$ case 
as indicated by the bound variation. This, in particular, means that this coupling ratio 
can be measured comparing the number of events in both channels. 
Obviously, the observed global strength of the $Z'$ couplings, $g'_{\rm L}$, depends on 
the total $Z'$ width, and the latter must be measured to determine the 
former. All these comments can be made quantitative running the corresponding 
Monte Carlo simulation but in this case it easy to prove these results analytically, too. 
Indeed, different final states do not interfere and in practice neither the two diagrams 
contributing to each process, because we require the reconstruction of the new gauge boson  
which is in general rather narrow. Hence, we can approximate the corresponding 
cross-sections assuming the narrow width approximation (which is good 
up to at most 20 \% when we compare to the numerical simulation): 
\begin{equation}
\sigma_{Z'}(pp\rightarrow X \mu^+ \mu^-) \approx \sigma_{X
  Z'}(M_{Z'}) {\rm Br}(Z'\rightarrow \mu^+ \mu^-) \ , 
\label{narrowwidthapproximation}
\end{equation}
and extrapolating the behavior of the diagram in Fig. \ref{zprimeproduction} we find: 
\begin{eqnarray}
&&\sigma_{Z'}(pp\rightarrow 3\mu + \cancel{E}_{\rm T}) \approx
f^{W}_{3\mu + \cancel{E}_{\rm T}}(M_{Z'})  
g^{\prime 2}_{\rm L} \frac{g^{\prime 2}_{\rm L} (1 + \xi^2)}{g^{\prime
    2}_{\rm L} (2 + \xi^2) + W^{i\neq \mu}} \ , \label{LHCcrosssections}   
\nonumber 
\\ 
&&\sigma_{Z'}(pp\rightarrow 4\mu) \approx f^{\gamma + Z}_{4\mu}(M_{Z'})
g^{\prime 2}_{\rm L} (2.5 + \xi^2)  
\frac{g^{\prime 2}_{\rm L} (1 + \xi^2)}{g^{\prime 2}_{\rm L} (2 +
  \xi^2) + W^{i\neq \mu}} \ . 
\end{eqnarray}
The fraction stands for the $Z'$ branching ratio (Br) into muons (see
Eqs. (\ref{zprimepartialwidth})  
and (\ref{totalwidth})), and the $Z'$ coupling dependence is derived
neglecting the  
EW gauge boson masses and assuming that the cross-section is dominated by the 
$\bar u u$ partonic contribution, as suggested by the proton content and the EW
couplings. 
We have numerically computed the exact factor without any of these 
approximations and found an excellent agreement, up to a negligible 
dependence on $m_{\mu^+ \mu^-}$\ .
We have also made an extra non-trivial check of the validity of Eqs.
(\ref{LHCcrosssections}) by plotting the exclusion limits reported 
in Fig. \ref{Mgxi} scaled by the 
$\xi$ dependence in Eqs. (\ref{LHCcrosssections}). The almost perfect
match is shown in Fig. \ref{Limitdependence}. The near equality of
$f^W_{3\mu +\cancel{E}_{\rm T}}(M_{Z^\prime})$ and
$f^{\gamma+Z}_{4\mu}(M_{Z^\prime})$ after selection cuts is, however, accidental.

\begin{figure}
\begin{centering}
\includegraphics[width=0.65\columnwidth]{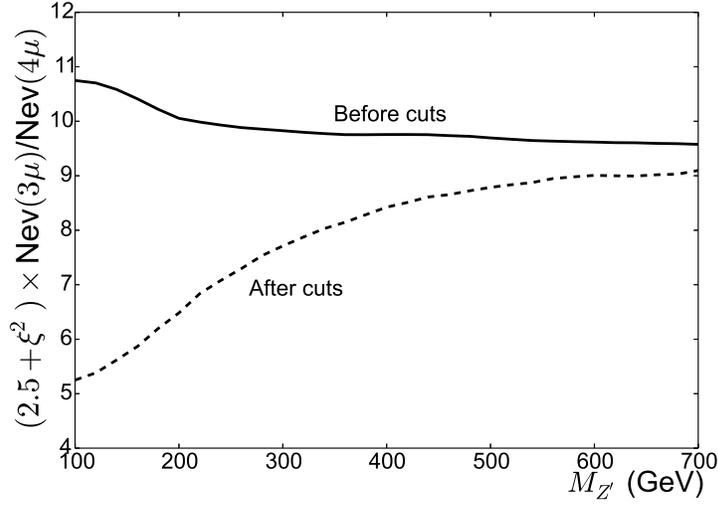}
\par\end{centering}
\caption{Ratio of 3$\mu$ plus missing energy to 4$\mu$ events as a function of the $Z'$ mass 
at the LHC. The quite different values before and after selection cuts are mainly due 
to the sensitivity to the cut on the invariant mass of the muon pair reconstructing the vector boson. 
\label{xidetermination}
}
\end{figure}

\begin{figure}
\begin{centering}
\includegraphics[width=0.65\columnwidth]{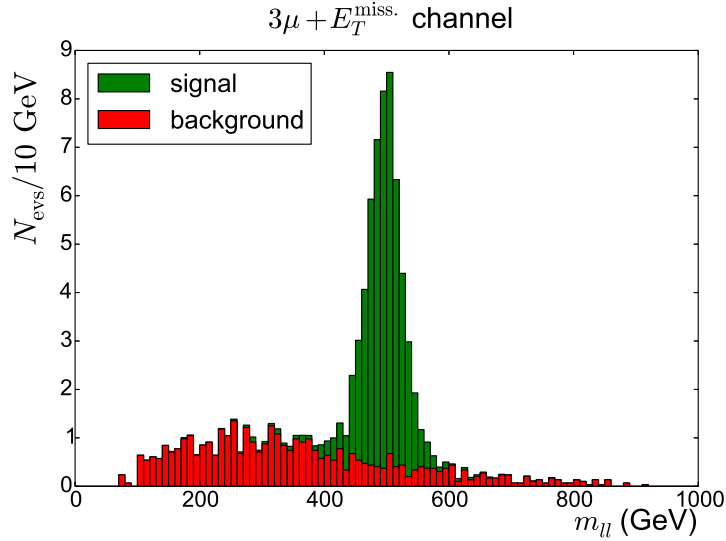}
\par\end{centering}
\caption{$Z'$ mass reconstruction in the $3\mu$ plus missing energy channel 
for $M_{Z'}=500$ GeV, $g'_{\rm L}=0.7$ and $\xi=1$, with $W^{i\neq \mu}=0$.
\label{LHCzprimewidth}
}
\end{figure}

Once we are convinced of the goodness of the quantitative analytic
results, we can attempt to  
determine $\xi$ from the number of events with 3$\mu$ plus missing
energy and 4$\mu$  
reconstructing a $Z' \rightarrow \mu^+  \mu^-$. 
This is done in Fig. \ref{xidetermination}, where we show the
corresponding cross-section ratio:  
\begin{equation}  
(2.5 + \xi^2) \ 
\frac{\sigma_{Z'}(pp\rightarrow 3\mu + \cancel{E}_{\rm T}^{\rm miss})}{\sigma_{Z'}(pp\rightarrow 4\mu)}  
\approx 9.4 \ , 
\label{RLratio}
\end{equation}
before and after cuts, as a function of the $Z'$ mass. 
Obviously, the approximation improves with the new vector boson mass and depends 
on the cuts, but it is clear that the ratio of the number of charge and neutral 
events is a sensitive probe of the value of $\xi$.
The determination of the $Z'$ width will also allow for the
measurement of the global strength of the $Z'$ couplings, $g'_{\rm L}$. 
This, however, does not seem to be easy at the LHC for the muon momentum 
measurement degrades for large values. In Fig. \ref{LHCzprimewidth} we plot 
the observed vector boson width 
\footnote{Experimentalists will certainly do better but this measurement is difficult 
due to the larger uncertainty associated to the determination of large muon momenta, 
as well as to the tendency of the $Z'$ to align with the emitting lepton and the relatively 
small number of signal events.}
$\sim 60$ GeV for a $Z'$ with a mass
of 500 GeV  
and a total width of 10 GeV.
Short of a precise measurement of the $Z'$ width we can still use the
fact that $W^{i\neq \mu}\geq 0$ to obtain a lower bound on $g'_{\rm L}$ 
from Eqs. (\ref{LHCcrosssections}). 
Other observables like, for instance, charged asymmetries can be also measured, 
but all the model dependence stays described by Eqs. (\ref{LHCcrosssections}). 

Given the relatively small cross-sections, the additional branching ratios and 
the missing energy accompanying any tau decay, 
it will be difficult to conclude anything using leptophilic final
states involving tau leptons.
Let us illustrate this in the anomaly-free muon minus tau LN case, ${\rm L}_\mu -
{\rm L}_\tau$,
\footnote{The limits in Fig. \ref{Mgxi} for $\xi = 1$ also apply to this
case but scaling the corresponding curve up by a factor $\sim \sqrt 2$,
to take into account for the $Z'_{\mu-\tau}$ total width ($W^{i \neq
\mu} \equiv W^\tau = W^\mu = g^{\prime 2}_{\rm L} (2 + \xi^2)$ in Eqs.
(\ref{totalwidth}) and (\ref{LHCcrosssections})). \label{footnote}}
for 2$\mu \tau_{\rm h} \nu_\tau$ events, where $\tau_{\rm h}$ stands for a
tau lepton decaying hadronically. 
Even in this mode with the handle of an opposite-sign muon
pair reconstructing the
$Z'_{\mu-\tau}$ mass, the SM backgrounds remain problematic.
For example, in this channel
the number of signal events is $\sim 51 \times g^{\prime 2}_{\rm L}$ for
$M_{Z'_{\mu-\tau}} = 300$ GeV and an integrated luminosity of 3 ab$^{-1}$. 
The number of $\bar t t$ events which passes our selection cuts is $\sim 228$,
where we have assumed that the probability for a jet to be misidentified as a
tau lepton decaying
hadronically is $10^{-3}$, which is rather conservative,
and a $k$-factor equal to 1.9 for $\bar t t$ production.
(Irreducible SM backgrounds with several EW gauge bosons are negligible.)
The applied cuts are the same as in Table \ref{3mu_cuts:table},
but with $p^{\tau_{\rm h}}_{\rm T} (> 20$ GeV) the lepton momentum
used to compute the transverse mass and $|\eta_{\tau_{\rm h}}| < 2.4$.
Just considering the $\bar t t$ background we need 31 (80) signal events to
set the 95 \% (5$\sigma$) exclusion (discovery) limit.
This translates into $g^{\prime}_{\rm L} < 0.78 \ (1.25)$ for
$M_{Z'_{\mu-\tau}} = 300$ GeV,
above the trident bound in both cases (see Fig \ref{Mgxi}).
Thus, we must rely on an experimental performance (analysis)
much better that this estimate to extract any information on leptophilic
final states with tau leptons.

\section{ILC reach for a leptophilic $Z'$}
\label{ILC}

As we will argue in the following, it will be eventually possible to measure 
the tau couplings of a leptophilic $Z'$ at the ILC, but not at the LHC 
as just illustrated in the previous section. In this sense, 
both machines are complementary, although no measurement of a 
leptophilic vector boson can be done at the ILC if such a $Z'$ is not previously 
observed at the LHC.  

The ILC production mechanism for a leptophilic vector boson with vanishing coupling to 
electrons is the same as at the LHC in Fig. \ref{zprimeproduction} but with the 
initial strong interacting partons replaced by an $e^+ e^-$ pair. 
The size of the cross-section is of the same order in both machines and 
the phenomenological search similar, too. 
However, at the ILC the initial state is neutral and only the neutral EW gauge 
bosons, $\gamma$ and $Z$, can be exchanged in the $s-$channel, although the 
three possible final states, 4$\mu$, 2$\mu$2$\nu_{\mu, \tau}$ and 2$\mu$2$\tau$, 
with a muon pair reconstructing the leptophilic 
vector boson, $Z' \rightarrow \mu^+ \mu^-$, emerge over the background in this case.
Moreover, using in the last two channels, 2$\mu$2$\nu_{\mu, \tau}$ and 2$\mu$2$\tau$, 
the two muon momenta and the total momentum, $P$, to reconstruct the new vector 
boson invariant mass, $M_{Z'}^2 = (P-p_{\mu^+}-p_{\mu^-})^2$, we can sample the 
four-lepton events with invisible $Z'$ decays, $Z' \rightarrow \bar \nu \nu$, in the first 
case and the four-lepton events with the new vector boson decaying into taus, $Z' \rightarrow \tau^+ \tau^-$, 
in the second one and hence, define two other differentiated sets of events (processes). 
Thus, we will deal with five different samples at the ILC, with the tau leptons identified by their 
hadronic decays, $\tau_{\rm h}$. 
The measurement of the corresponding cross-sections could, a priori, overdetermine the 
$Z'$ couplings to LH and RH muons, $g^{\prime}_{\rm L}$ and $\xi g^{\prime}_{\rm L}$, 
as well as to LH and RH taus, $g^{\prime \tau}_{\rm L}$ and $\xi_\tau g^{\prime \tau}_{\rm L}$, 
once the total $Z'$ width (Eq. (\ref{totalwidth})) is determined fitting the corresponding 
Breit-Wigner distribution to the lepton pairs reconstructing the $Z'$ mass. 
However, in practice, as we shall make explicit below, the similar coupling 
dependence of the two 2$\mu$2$\nu_{\mu, \tau}$ subsamples on one hand and of the two 
2$\mu$2$\tau$ ones on the other, which is accidental and a consequence of 
the particular value of the SM mixing angle, only allows for the determination of 
three coupling constant combinations. 
But the two tau couplings enter in two of them and with different 
dependence and hence, even with large errors due to the small 
cross-sections involved, both can be determined at the ILC. 
In any case, the muon couplings can be more precisely measured at the LHC, but 
not the $Z'$ total width for low vector boson masses which are the only accessible at the ILC.

Applying the same approximations as for the LHC in the former section, 
we can also extrapolate the corresponding cross-sections at the ILC, obtaining:
\begin{eqnarray}
\label{ILCcrosssections}
\sigma_{Z'\rightarrow \mu^+ \mu^-}(e^+ e^- \rightarrow 4\mu) \approx 
f^{\prime \gamma + Z}_{4\mu}(M_{Z'}) g^{\prime 2}_{\rm L} (1.15 + \xi^2) 
\frac{g^{\prime 2}_{\rm L} (1 + \xi^2)}{g^{\prime 2}_{\rm L} (2 + \xi^2) + W^{i\neq \mu}} \ , 
\quad \quad \quad \quad \nonumber \\ 
\sigma_{Z'\rightarrow \mu^+ \mu^-}(e^+ e^- \rightarrow 2\mu + \cancel{E}_{\rm T}) \approx 
f^{\prime Z}_{2\mu + \cancel{E}_{\rm T}}(M_{Z'}) 
(g^{\prime 2}_{\rm L} + g^{\prime \tau 2}_{\rm L}) 
\frac{g^{\prime 2}_{\rm L} (1 + \xi^2)}{g^{\prime 2}_{\rm L} (2 + \xi^2) + W^{i\neq \mu}} \ , \quad \quad \nonumber \\ 
\sigma_{Z'\rightarrow \bar \nu \nu}(e^+ e^- \rightarrow 2\mu + \cancel{E}_{\rm T}) \approx 
f^{\prime \gamma + Z}_{2\mu + \cancel{E}_{\rm T}}(M_{Z'}) 
g^{\prime 2}_{\rm L} (1.15 + \xi^2) 
\frac{g^{\prime 2}_{\rm L} + g^{\prime \tau 2}_{\rm L}}{g^{\prime 2}_{\rm L} (2 + \xi^2) + W^{i\neq \mu}} 
\ , \quad \quad \\ 
\sigma_{Z'\rightarrow \mu^+ \mu^-}(e^+ e^- \rightarrow 2\mu + 2\tau) \approx 
f^{\prime \gamma + Z}_{2\mu + 2\tau}(M_{Z'})
g^{\prime \tau 2}_{\rm L} (1.15 + \xi^{2}_\tau) 
\frac{g^{\prime 2}_{\rm L} (1 + \xi^2)}{g^{\prime 2}_{\rm L} (2 + \xi^2) + W^{i\neq \mu}} \ , 
\; \; \; \; \nonumber \\
\sigma_{Z'\rightarrow \tau^+ \tau^-}(e^+ e^- \rightarrow 2\mu + 2\tau) \approx 
f^{\prime \gamma + Z}_{2\mu + 2\tau}(M_{Z'})
g^{\prime 2}_{\rm L} (1.15 + \xi^2)
\frac{g^{\prime \tau 2}_{\rm L} (1 + \xi^{2}_\tau)}{g^{\prime 2}_{\rm L} (2 + \xi^2) + W^{i\neq \mu}} \ , \; \; \; 
\quad \nonumber 
\end{eqnarray}
where if the tau couplings are non-vanishing, $W^{i\neq \mu}$ includes at least the tau contribution 
$g^{\prime \tau 2}_{\rm L} (2 + \xi_\tau^2)$. 
As pointed out, the second and third equations have a very similar 
coupling constant dependence due to the particular value of the Weinberg angle, 
which results in the term 1.15 in the third cross-section to be compared with the 
unit term in the branching ratio in the second one; 
and analogously for the fourth and fifth equations. 
Although the muon couplings can be measured at the LHC, both tau couplings can be 
only determined at the ILC, providing a new example of the complementarity of both colliders. 
After cuts all (sub)samples will differ, except for the last two which we will analyze (add) together. 
Before selection cuts, however, 
\begin{equation}
f^{\prime \gamma + Z}_{4\mu}(M_{Z'}) \approx 2  f^{\prime Z}_{2\mu + \cancel{E}_{\rm T}}(M_{Z'}) 
\approx f^{\prime \gamma + Z}_{2\mu + \cancel{E}_{\rm T}}(M_{Z'}) \approx f^{\prime \gamma + Z}_{2\mu + 2\tau}(M_{Z'})\ .
\label{scales}
\end{equation}

In order to estimate the ILC reach for different samples we have followed
the same generation procedure as for the LHC. 
Hadronic taus, $\tau_{\rm h}$, are tagged by a pure geometrical method, becoming a jet a potential hadronic tau 
if a generated tau is found within a fixed distance $\Delta R=0.5$ of the jet axis, with an efficiency of 0.5. 
For each particular sample, we impose a different set of cuts in order to isolate the signal from the background. 
They also allow for discriminating between the different signal samples. 
The cuts are shown in Table \ref{ILCcuts}, and have been implemented 
using \texttt{MADANALYSIS 5}~\cite{Conte:2012fm}. 
From top to bottom, they refer to the samples $4\mu$, $2\mu 2\nu_{\mu,\tau}$ 
with $Z'$ decaying into muons, $2\mu 2\nu_{\mu,\tau}$ with $Z'$ decaying into neutrinos and $2\mu2\tau$ 
with $Z'$ decaying into both muons and taus, 
with at least one tau lepton decaying hadronically. 
In the last case we sum both subsamples to improve the statistics because their model dependence 
and the efficiency in the sampling are very similar. 
$m_{\bar{\nu}\nu}$ and $m_{\tau^+\tau^-}$ stand for the invariant mass reconstructed from the two 
observed muons and the initial momentum, $\sqrt{(P - p_{\mu^+} - p_{\mu^-})^2}$. 
\begin{table}[]
\begin{center}
\begin{tabular}{lcl}\toprule
Basic cuts & \hspace{1cm} & $p_T^\ell > 10$ GeV, $|\eta_\ell| < 2.47$, $\Delta R(j\ell) > 0.4$, 
$p_T^j > 20$ GeV, $|\eta_j| < 2.5$ \\\midrule
Number of muons & & $N_\mu = 4$ (zero net charge) \\
$Z$ veto & & $|m_{\mu^+\mu^-} - M_Z| > 10$ GeV\\
Mass window & & $|m_{\mu^+\mu^-} -M_{Z'}| < 10$ GeV\\\midrule
Number of muons & & $N_\mu = 2$ (zero net charge) \\
Number of taus & & $N_{\tau_{\rm h}} = 0$\\
$Z$ veto & & $|m_{\mu^+\mu^-} - M_Z|$ and $|m_{\bar{\nu}\nu} - M_Z| > 10$ GeV\\
Mass window & & $|m_{\mu^+\mu^-} -M_{Z'}| < 10$ GeV\\\midrule
Number of muons & & $N_\mu = 2$ (zero net charge) \\
Number of taus & & $N_{\tau_{\rm h}} = 0$\\
$Z$ veto & & $|m_{\mu^+\mu^-} - M_Z|$ and $|m_{\bar{\nu}\nu} - M_Z| > 10$ GeV\\
Mass window & & $|m_{\bar{\nu}\nu} - M_{Z'}| < 10$ GeV\\\midrule
Number of muons & & $N_\mu = 2$ (zero net charge) \\
Number of taus & & $N_{\tau_{\rm h}} \geq 1$\\
$Z$ veto & & $|m_{\mu^+\mu^-} - M_Z| $ and $|m_{\tau^+\tau^-} - M_Z| > 10$ GeV\\
Mass window & & $|m_{\mu^+\mu^-} - M_{Z'}|$ or $|m_{\tau^+\tau^-} - M_{Z'}| < 10$ GeV\\\midrule
\end{tabular}
\caption{From top to bottom, cuts imposed on the samples $4\mu$, $2\mu 2\nu_{\mu,\tau}$ 
with $Z'$ decaying into muons, $2\mu 2\nu_{\mu,\tau}$ with $Z'$ decaying into neutrinos and $2\mu2\tau$ 
with $Z'$ decaying into both muons and taus, respectively. 
$m_{\bar{\nu}\nu}$ and $m_{\tau^+\tau^-}$ stand for the invariant mass reconstructed from the two 
observed muons and the initial total momentum, $\sqrt{(P - p_{\mu^+} - p_{\mu^-})^2}$. 
\label{ILCcuts}}
\end{center}
\end{table}

Only irreducible backgrounds are considered for each case. We have checked that other backgrounds are negligible after applying the cuts in Table \ref{ILCcuts}. 
In particular, in the $2\mu 2\tau$ case $Z+$jets is subleading given the small fake-rate for tau tagging (of around $10^{-3}$).

\begin{figure}
\begin{center}
\includegraphics[width=0.49\columnwidth]{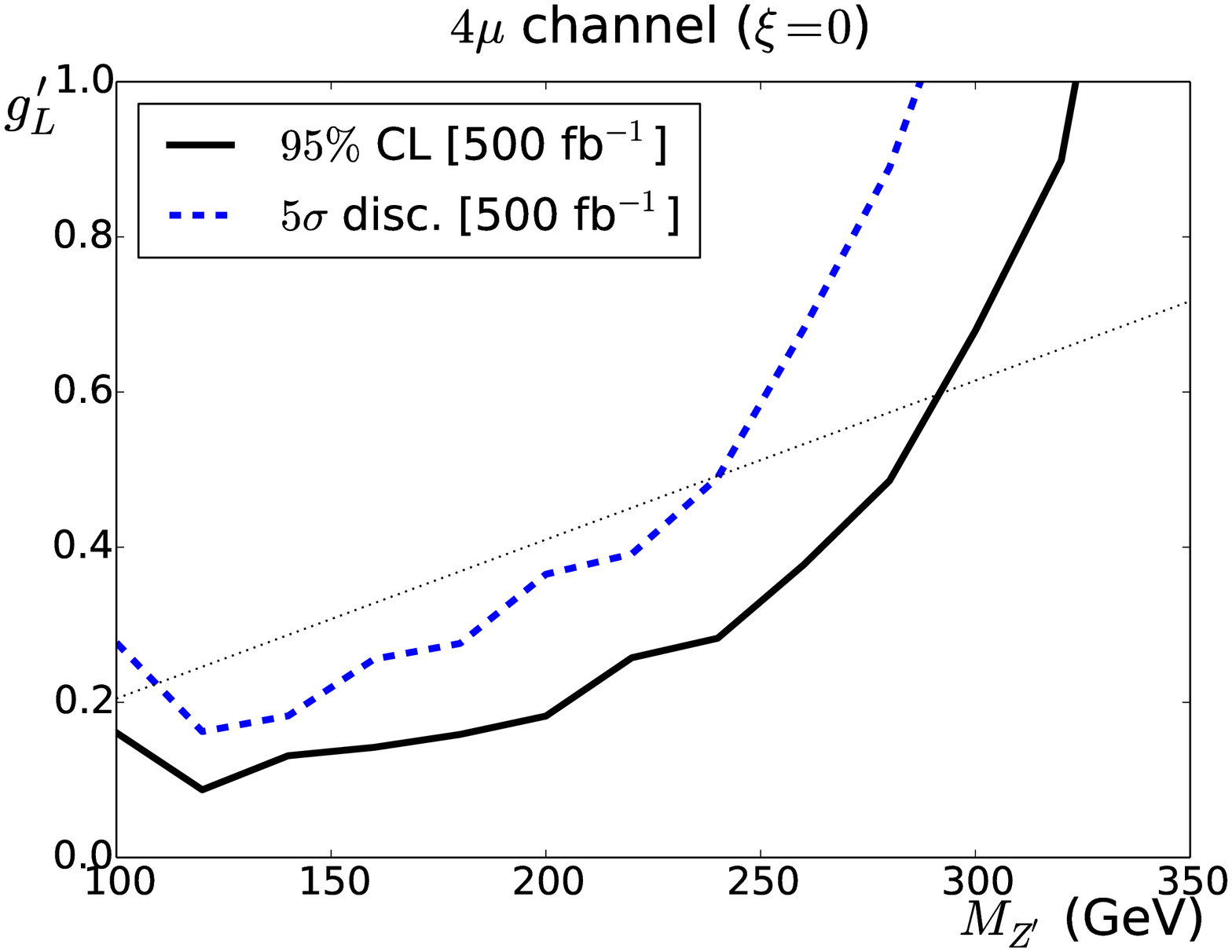}
\includegraphics[width=0.49\columnwidth]{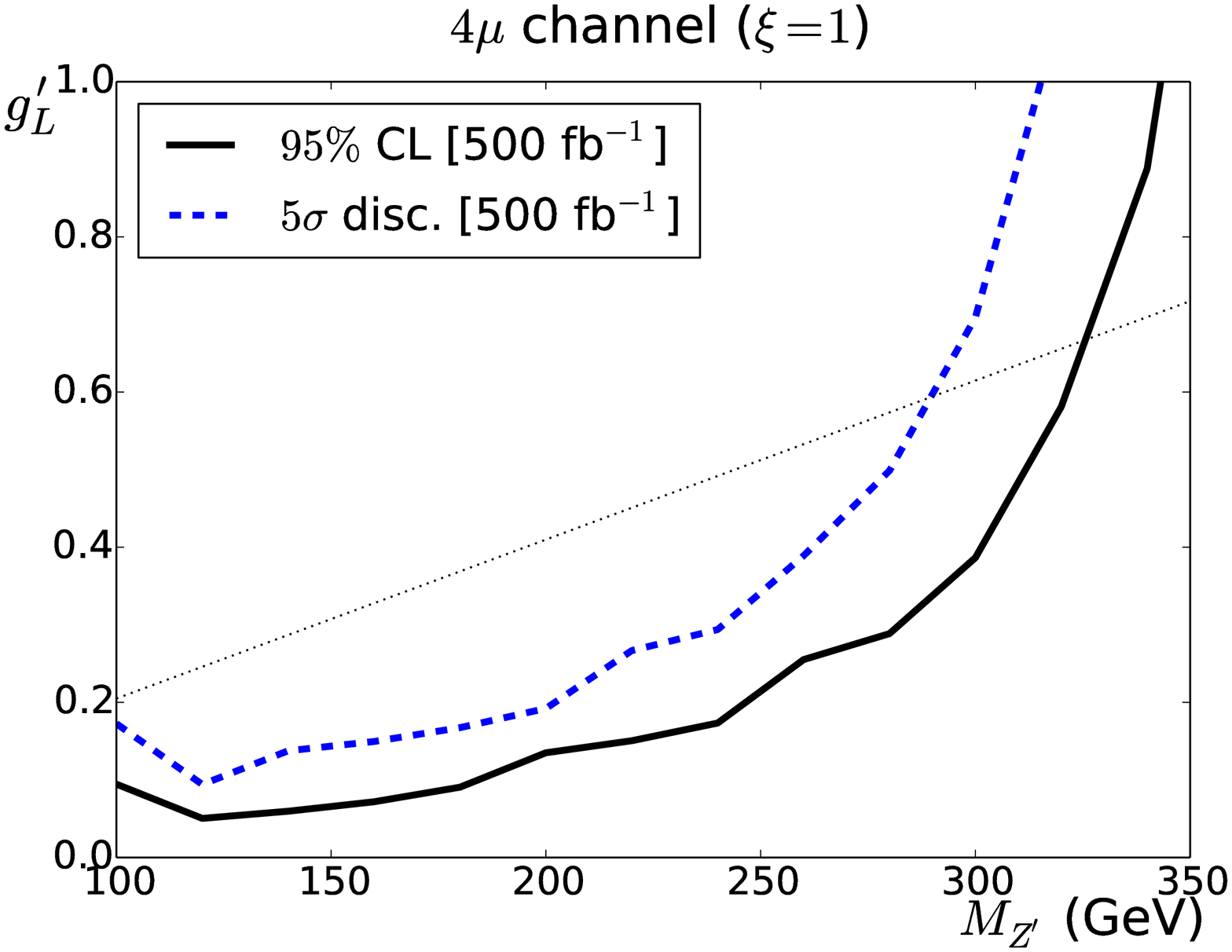}
\par\end{center}
\caption{Discovery and exclusion limits for a leptophilic $Z'$ as a function of 
its mass and its coupling to LH muons for $\xi = 0$ (left) and $\xi = 1$ (right) 
using only 4$\mu$ events at the ILC. We assume that no other $Z'$ decay channel is open (in particular, 
that $Z'$ does not decay into taus). The bound from neutrino trident
production (straight line) is also shown for comparison.
\label{ILCMgxi}
}
\end{figure}

Analogously as for the LHC, in Fig. \ref{ILCMgxi} we plot the ILC discovery (5$\sigma$) and exclusion 
(95 \% C.L.) limits for a leptophilic $Z'$ as a function of its mass and coupling to LH muons 
using only 4$\mu$ events. 
($Z'$ couplings to taus are neglected if not stated otherwise.) 
The bounds from neutrino trident production 
are also shown, 
for comparison (see the former section for details) 
\footnote{The limits from $g-2$ are less stringent when comparable 
(see Ref. \cite{Harigaya:2013twa,Bell:2014tta,Ma:2001md,Altmannshofer:2014cfa}).}. 
Obviously, although the ILC limits are similar to 
the LHC ones for low $Z'$ masses, the bounds rapidly deteriorate 
for $Z'$ masses near the ILC CME (compare Figs. \ref{Mgxidiscovery} and 
\ref{Mgxi} with \ref{ILCMgxi}). 

\begin{figure}
\begin{centering}
\includegraphics[width=0.65\columnwidth]{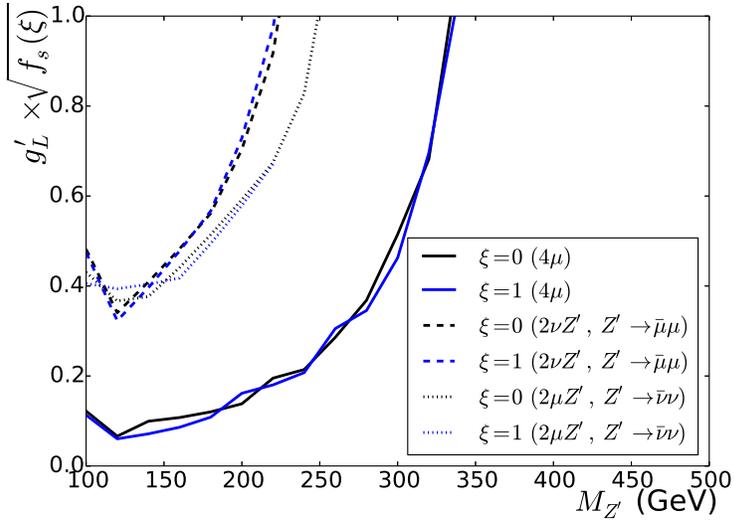}
\par\end{centering}
\caption{95 \% C.L. exclusion limits for a leptophilic $Z'$ as a function of 
its mass and its coupling to LH muons for two values of its coupling 
to their RH counterparts at the ILC, corrected for the new vector boson coupling dependence 
in Eqs. (\ref{ILCcrosssections}): 
$f_{4\mu}(\xi) = (1.15 + \xi^2) \frac{1 + \xi^2}{2 + \xi^2}$, 
$f_{2\mu}(\xi) = \frac{1 + \xi^2}{2 + \xi^2}$ for $Z' \rightarrow \mu^+ \mu^-$ and
$f_{2\mu}(\xi) = \frac{1.15 + \xi^2}{2 + \xi^2}$ for $Z' \rightarrow \bar \nu \nu$. 
We assume that the new gauge boson does not 
couple (decay) to tau leptons.
\label{ILCLimitdependence}
}
\end{figure}

\begin{figure}
\begin{center}
\includegraphics[width=0.49\columnwidth]{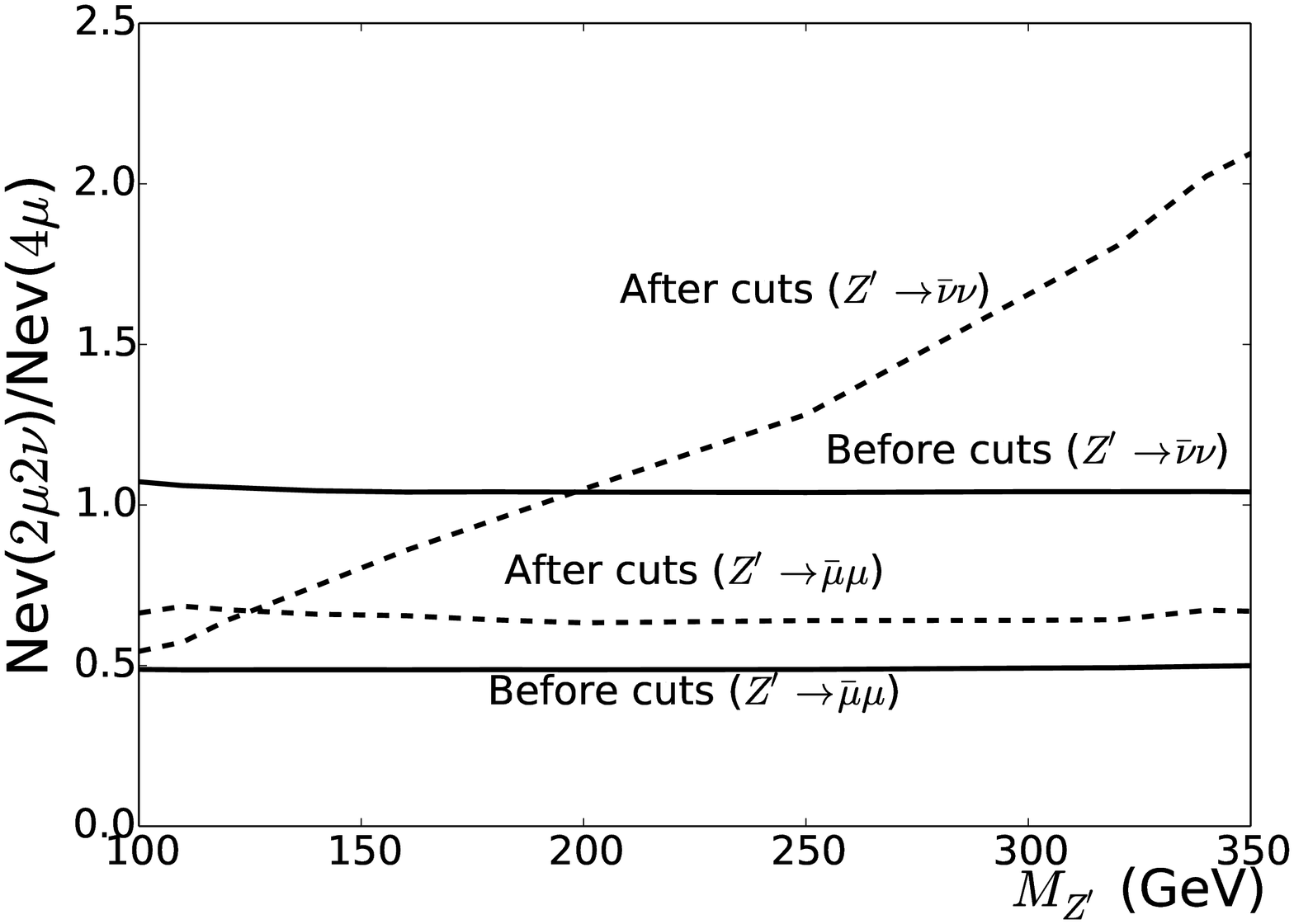}
\includegraphics[width=0.49\columnwidth]{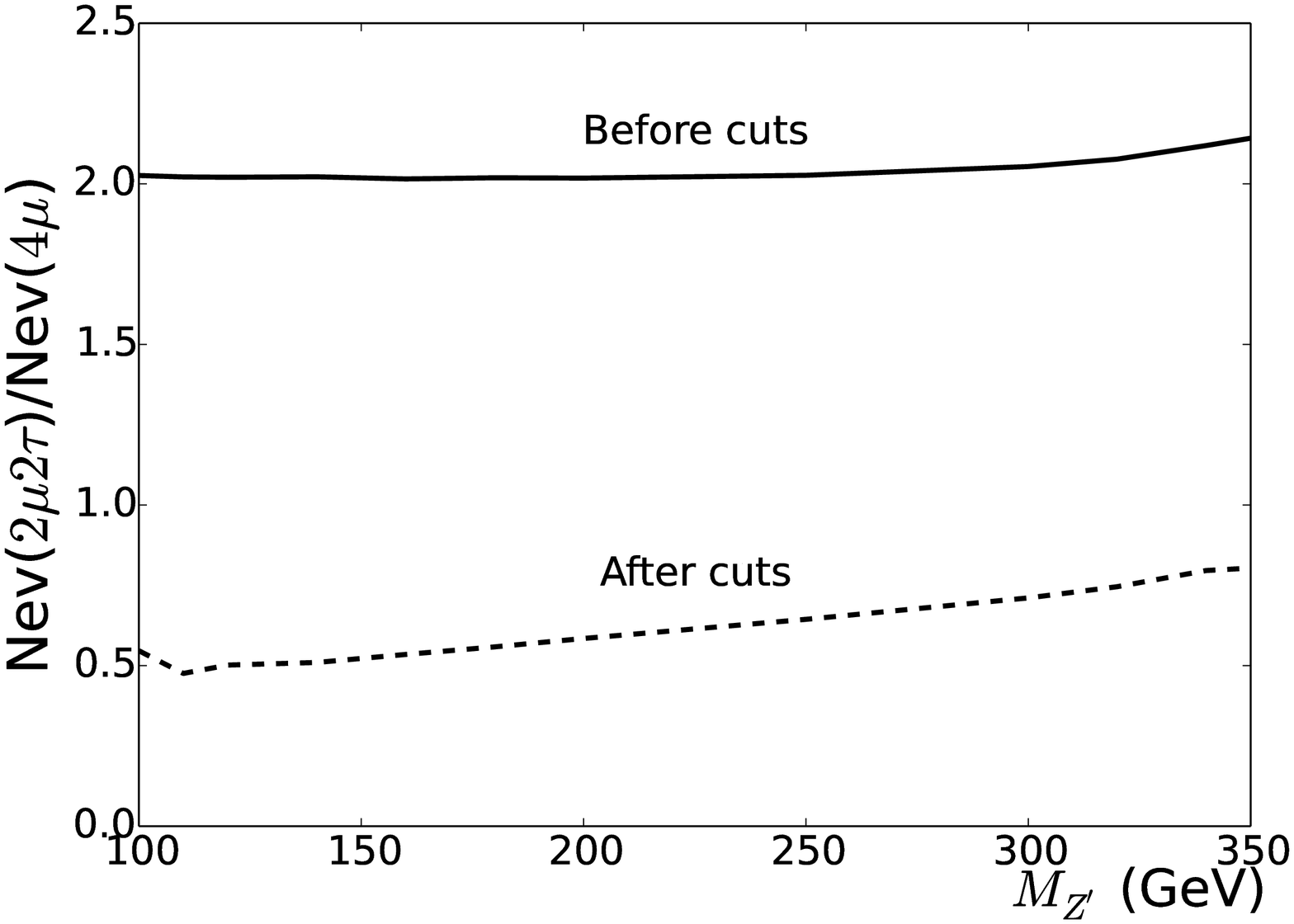}
\end{center}
\caption{(Left) Ratio of 2$\mu$ plus missing energy to 4$\mu$ events 
as a function of the $Z'_{\mu - \tau}$ mass for the anomaly-free model gauging 
${\rm L}_\mu - {\rm L}_\tau$ at the ILC, distinguishing both vector boson decay modes 
$Z'_{\mu - \tau} \rightarrow \mu^+\mu^-$ and $\bar \nu \nu$. 
(Right) The same but for the ratio of 2$\mu$2$\tau$ to 4$\mu$ events. 
\label{couplingdetermination}
}
\end{figure}

Also similarly as for the LHC, we want to test how good the 
approximations in Eqs. (\ref{ILCcrosssections}) are, and how well 
the $Z'$ properties can be determined. 
Thus, in Fig. \ref{ILCLimitdependence} we draw, for instance, the exclusion limits 
for each four-lepton subsample scaled by the coupling dependence 
in Eqs. (\ref{ILCcrosssections}) 
\footnote{As the $Z'$ is assumed not to couple to taus in this case, only the first three 
processes and equations are relevant.}, 
obtaining again an almost perfect matching.  
To measure the tau couplings we must also confront the 2$\mu$2$\tau$ sample. 
Hence, in Fig. \ref{couplingdetermination} we plot 
for the ${\rm L}_\mu - {\rm L}_\tau$ model 
the ratios of 2$\mu$ plus missing energy to 4$\mu$ events (left), distinguishing 
both $Z'_{\mu - \tau}$ decays to $\mu^+ \mu^-$ and $\bar \nu \nu$, and of 
2$\mu$2$\tau$ to 4$\mu$ events (right) as a function of the $Z'_{\mu - \tau}$ mass. 
As can be observed, the curves before selection cuts closely follow
the proportionality relations in  
Eq. (\ref{scales}). What is not the case after selection cuts mainly
due to the large sensitivity of some of the samples to the different cuts,
especially to the fixed Mass window bin.

\begin{figure}
\begin{centering}
\includegraphics[width=0.65\columnwidth]{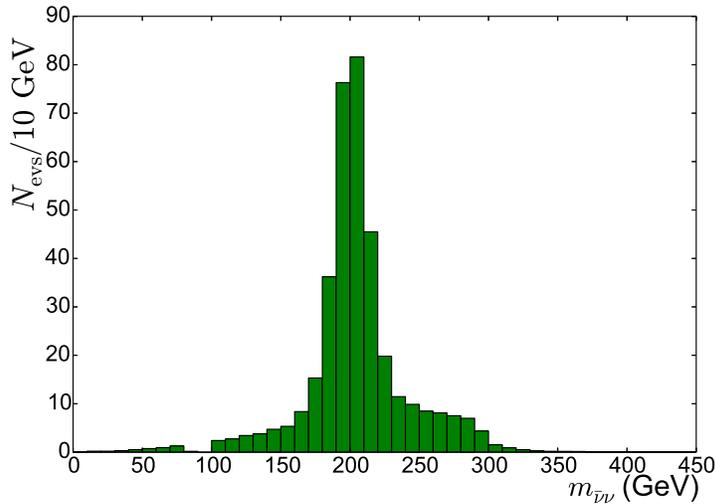}
\par\end{centering}
\caption{$Z'_{\mu - \tau} \rightarrow \bar \nu \nu$ mass reconstruction in the $2\mu 2\nu$ channel 
for $M_{Z'}=200$ GeV and $g'_{\rm L}=1$.
\label{ILCzprimewidth}
}
\end{figure}

Finally, the total $Z'$ width can be eventually measured at the ILC for a large enough statistics, 
but in this case the accessible $Z'$ masses are lower than at the LHC (and its alignment with the 
emitting lepton is less pronounced), as there are lower the muon momenta and 
then smaller the uncertainty in their determination. 
As already stated, the $Z'$ width can be measured in two ways 
depending on whether we reconstruct the vector boson mass with two muons or 
two neutrinos, being the second one, $M^2 _{Z'} = (P - p_{\mu^+} - p_{\mu^-})^2$, 
a priori more precise. In Fig. \ref{ILCzprimewidth} we show the 
observed width $\sim 30$ GeV for a $Z'$ of 200 GeV and a total width of 16 GeV, 
assuming that the detector performance is the same as for the LHC, although it is 
aimed to be better.

In summary, once measured the total $Z'$ width, and determined the muon couplings at the LHC, 
$g^{\prime \tau}_{\rm L}$ and $\xi_\tau$ can be measured at the ILC 
for low $Z'$ masses, but most probably with a large statistical error. 
We have also investigated the possibility of improving the analysis 
of the 2$\mu$2$\tau$ sample by requiring not one (see the selection cuts in Table \ref{ILCcuts})  
but both tau leptons to decay hadronically. However, although similar, the derived limits are less 
stringent and the tau coupling determination less precise for the heavier part of the spectrum 
due to the extra branching ratio suppression.

\section{Large collider bounds on four-lepton invariant operators}
\label{4loperators}

\begin{figure}
\begin{centering}
\includegraphics[width=0.55\columnwidth]{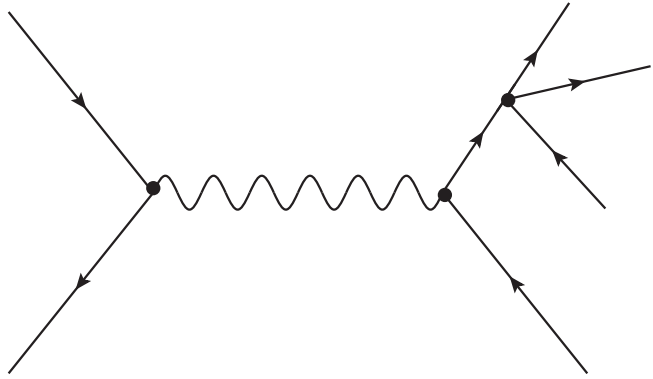}
\par\end{centering}
\caption{Leading contribution of four-lepton interactions not coupling to electrons at the LHC and ILC.
\label{fourleptonproduction}
}
\end{figure}

The large collider limits on an extra vector boson only coupling to 
muons and taus, which we derived in previous sections, 
were obtained assuming a relatively narrow resonance and hence, an event excess 
around the vector boson mass. 
The non-observation of such an excess not only for any lepton-pair invariant 
mass but for the appropriate integrated region of four-lepton events can be also used to bound the 
size of the tail of new leptophilic interactions. 
In general these can be parametrized by the corresponding 
four-lepton operators, which contribute to four-lepton samples through the diagram(s) in 
Fig. \ref{fourleptonproduction}. 
The cuts in this case, however, are different, as there are the bounds 
and the physical interpretation. 

Let us first study the case of a $Z'_{\mu - \tau}$ somewhat heavier than 
the LHC reach for illustration. 
The integration of this gauge boson out generates the four-lepton effective 
Lagrangian \cite{delAguila:2010mx}: 
\begin{equation}
\begin{tabular}{ll} 
${\cal L}_{\rm eff} = - \frac{g_{\rm L}^{\prime 2}}{2M_{Z'_{\mu-\tau}}^2}$ 
& $[(\overline{L_{{\rm L} \mu}} \gamma_\alpha L_{{\rm L} \mu})(\overline{L_{{\rm L} \mu}} \gamma^\alpha L_{{\rm L} \mu}) 
- 2 (\overline{L_{{\rm L} \mu}} \gamma_\alpha L_{{\rm L} \mu})(\overline{L_{{\rm L} \tau}} \gamma^\alpha L_{{\rm L} \tau})$ \\ 
& + $(\overline{L_{{\rm L} \tau}} \gamma_\alpha L_{{\rm L} \tau})(\overline{L_{{\rm L} \tau}} \gamma^\alpha L_{{\rm L} \tau}) 
+ (\overline{\mu_{\rm R}} \gamma_\alpha \mu_{\rm R})(\overline{\mu_{\rm R}} \gamma^\alpha \mu_{\rm R})$ \\
& $-\ 2 (\overline{\mu_{\rm R}} \gamma_\alpha \mu_{\rm R})(\overline{\tau_{\rm R}} \gamma^\alpha \tau_{\rm R}) 
+ (\overline{\tau_{\rm R}} \gamma_\alpha \tau_{\rm R})(\overline{\tau_{\rm R}} \gamma^\alpha \tau_{\rm R})$ \\ 
& $-\ 4 (\overline{L_{{\rm L} \mu}} \mu_{\rm R})(\overline{\mu_{\rm R}} L_{{\rm L} \mu}) 
+ 4 (\overline{L_{{\rm L} \mu}} \tau_{\rm R})(\overline{\tau_{\rm R}} L_{{\rm L} \mu})$ \\ 
& + $4 (\overline{L_{{\rm L} \tau}} \mu_{\rm R})(\overline{\mu_{\rm R}} L_{{\rm L} \tau})
- 4 (\overline{L_{{\rm L} \tau}} \tau_{\rm R})(\overline{\tau_{\rm R}} L_{{\rm L} \tau})] ,$
\end{tabular}
\label{Zprimeintegration}
\end{equation}
which describes its low energy effects, and eventually the departure 
from the SM predictions for $4\mu$ and $3\mu$ plus missing
energy distributions at the LHC.  
In order to confront this particular SM extension with an eventual
excess or deficit of events in  
these samples, we have generated events as in previous sections but
implementing  
the effective Lagrangian in Eq. (\ref{Zprimeintegration}) in the \texttt{UFO} 
format~\cite{Degrande:2011ua}. 
We have then performed the corresponding analysis using also
\texttt{MADANALYSIS 5}~\cite{Conte:2012fm}.  
The cuts are the same as the ones in Tables~\ref{3mu_cuts:table} and
\ref{4mu_cuts:table} except 
for the last one (Mass window) which is replaced by the corresponding
cut in Table
\ref{table:Cutsoperators}.  
\begin{table}[]
\begin{center}
\begin{tabular}{lcll}
\toprule
Scalar sum & & $S_{\rm T} = \sum\limits_{i=1}^3 |p^{\mu_i}_{\rm T}| +
\cancel{E}_{\rm T} > 600$ GeV, &\quad  $3\mu$ plus missing energy channel
\\\midrule
Scalar sum & & $S_{\rm T} = \sum\limits_{i=1}^4 |p^{\mu_i}_{\rm T}| >
400$ GeV, & \quad $4\mu$ channel 
\\ 
\bottomrule
\end{tabular}
\caption{\label{table:Cutsoperators}
Further cut which replaces the last cut (Mass window) in
Tables~\ref{3mu_cuts:table} and \ref{4mu_cuts:table} 
for the 3$\mu$ plus missing energy and 4$\mu$ channels, respectively,
in the absence of light resonances.}
\end{center}
\end{table}
This does not require the vector boson reconstruction 
but integrates the events on the tail of the kinematical distribution
of the scalar transverse energy sum $S_{\rm T}$.  

Assuming that no departure from the SM predictions is observed, we 
obtain an upper bound on the effective Lagrangian coefficient 
$x = \frac{g_{\rm L}^{\prime 2}}{2M_{Z'_{\mu-\tau}}^2} < 3.8$ TeV$^{-2}$ at 95 \% C.L., 
where we have closely followed the approach in Ref. \cite{deBlas:2013qqa} to parametrize 
the effect of the four-lepton operators. 
This can be formally expressed as a limit on 
$M_{Z'_{\mu-\tau}} > 0.36\  g_{\rm L}^{\prime}$ TeV, too. 
But this region of parameter space is mainly excluded 
by neutrino trident production, and by the analysis presented in
Section~\ref{4m}
if no weakly coupled 
($g'_{\rm L} < \sqrt{4\pi}$) resonance ($Z'_{\mu-\tau}$) is found at
the LHC (see Fig. \ref{Mgxi} and footnote \ref{footnote}). 
Note that, on the other hand, in this particular case 
the limit on $x$ can be neither translated into a bound on $M_{Z'_{\mu-\tau}}$ 
using its explicit dependence and hence, can not be
interpreted as  
the limit on the effective coupling of a heavy resonance banished to a higher 
energy scale. This is always the case when 
the allowed values of $g_{\rm L}^{\prime}$ exceed $\sqrt{4\pi}$
for masses not already excluded by the search of the corresponding weakly coupled 
resonances (at the LHC in our case).  

Although in general this will be also the case for the LHC and ILC limits 
derived for any arbitrary combination of four-lepton operators 
involving muons and taus only, we can perform the same analysis as for 
${\cal L}_{\rm eff}$ in Eq. (\ref{Zprimeintegration}) 
and obtain bounds on the corresponding effective Lagrangian coefficients. 
With this purpose in the following we define a basis of such four-lepton Lorentz 
and gauge invariant operators preserving flavor parity, and derive limits 
for the most favorable cases. 
Thus, in Table \ref{mutaufourleptonoperators} we list such an operator basis, 
up to hermitian conjugation.
\begin{table}[]
\begin{center}
{
\begin{tabular}{ rlc|crl } 
\multicolumn{2}{c}{$LL$ and $ll$ perators} & \quad\quad & \quad\quad & \multicolumn{2}{c}{$Ll$ operators} \\
\hline
$({\cal O}^{(1)}_{LL})_{\mu\mu\mu\mu} =$ & $\frac{1}{2} (\overline{L_{{\rm L} \mu}} \gamma_\alpha L_{{\rm L} \mu})
(\overline{L_{{\rm L} \mu}} \gamma^\alpha L_{{\rm L} \mu})$ 
& & & $({\cal O}_{Ll})_{\mu\mu\mu\mu} =$ & $(\overline{L_{{\rm L} \mu}} \mu_{\rm R})(\overline{\mu_{\rm R}} L_{{\rm L} \mu}) $ \\ 
$({\cal O}^{(1)}_{LL})_{\mu\mu\tau\tau} =$ & $\frac{1}{2} (\overline{L_{{\rm L}\mu}} \gamma_\alpha L_{{\rm L} \mu})
(\overline{L_{{\rm L} \tau}} \gamma^\alpha L_{{\rm L} \tau})$ 
& & & $({\cal O}_{Ll})_{\mu\mu\tau\tau} =$ & $(\overline{L_{{\rm L} \mu}} \mu_{\rm R})(\overline{\tau_{\rm R}} L_{{\rm L} \tau}) $ \\ 
$({\cal O}^{(1)}_{LL})_{\mu\tau\mu\tau} =$ & $\frac{1}{2} (\overline{L_{{\rm L} \mu}} \gamma_\alpha L_{{\rm L} \tau})
(\overline{L_{{\rm L} \mu}} \gamma^\alpha L_{{\rm L} \tau})$ 
& & & $({\cal O}_{Ll})_{\mu\tau\mu\tau} =$ & $(\overline{L_{{\rm L} \mu}} \tau_{\rm R})(\overline{\mu_{\rm R}} L_{{\rm L} \tau}) $ \\ 
$({\cal O}^{(1)}_{LL})_{\tau\tau\tau\tau} =$ & $\frac{1}{2} (\overline{L_{{\rm L} \tau}} \gamma_\alpha L_{{\rm L} \tau})
(\overline{L_{{\rm L} \tau}} \gamma^\alpha L_{{\rm L} \tau})$ 
& & & $({\cal O}_{Ll})_{\mu\tau\tau\mu} =$ & $(\overline{L_{{\rm L} \mu}} \tau_{\rm R})(\overline{\tau_{\rm R}} L_{{\rm L} \mu})$ \\ 
$({\cal O}_{ll})_{\mu\mu\mu\mu} =$ & $\frac{1}{2} (\overline{\mu_{\rm R}} \gamma_\alpha \mu_{\rm R})
(\overline{\mu_{\rm R}} \gamma^\alpha \mu_{\rm R})$ 
& & & $({\cal O}_{Ll})_{\tau\mu\mu\tau} =$ & $(\overline{L_{{\rm L} \tau}} \mu_{\rm R})(\overline{\mu_{\rm R}} L_{{\rm L} \tau})$ \\ 
$({\cal O}_{ll})_{\mu\mu\tau\tau} =$ & $\frac{1}{2} (\overline{\mu_{\rm R}} \gamma_\alpha \mu_{\rm R})
(\overline{\tau_{\rm R}} \gamma^\alpha \tau_{\rm R})$ 
& & & $({\cal O}_{Ll})_{\tau\tau\tau\tau} =$ & $(\overline{L_{{\rm L} \tau}} \tau_{\rm R})(\overline{\tau_{\rm R}} L_{{\rm L} \tau})$ \\ 
$({\cal O}_{ll})_{\mu\tau\mu\tau} =$ & $\frac{1}{2} (\overline{\mu_{\rm R}} \gamma_\alpha \tau_{\rm R})
(\overline{\mu_{\rm R}} \gamma^\alpha \tau_{\rm R})$ 
& & & & \\ 
$({\cal O}_{ll})_{\tau\tau\tau\tau} =$ & $\frac{1}{2} (\overline{\tau_{\rm R}} \gamma_\alpha \tau_{\rm R})
(\overline{\tau_{\rm R}} \gamma^\alpha \tau_{\rm R})$ 
& & & & \\ 
\end{tabular}
}
\caption{\label{mutaufourleptonoperators} 
Independent four-lepton (gauge invariant) operators ${\cal O}_{4l}$ involving muons and taus only. 
We assume flavor-parity conservation and omit hermitian conjugated partners.}
\end{center}
\end{table}
That this set is complete can be shown using the Fierz relation
\begin{equation}
({\cal O}_{LL}^{(3)})_{ijkl} = 2 ({\cal O}_{LL}^{(1)})_{ilkj} - ({\cal O}_{LL}^{(1)})_{ijkl}\ ,
\end{equation}
to reduce the four combinations of four SM lepton multiplets which are Lorentz and gauge invariant: 
\begin{equation}
\begin{tabular}{ rl } 
$({\cal O}_{LL}^{(1)})_{ijkl} =$ & $\frac{1}{2} (\overline{L_{{\rm L} i}} \gamma_\alpha L_{{\rm L} j})
(\overline{L_{{\rm L} k}} \gamma^\alpha L_{{\rm L} l})$ , 
 \\ 
$({\cal O}_{LL}^{(3)})_{ijkl} =$ & $\frac{1}{2} (\overline{L_{{\rm L} i}} \gamma_\alpha \sigma_a L_{{\rm L} j})
(\overline{L_{{\rm L} k}} \gamma^\alpha \sigma_a L_{{\rm L} l})$ , 
\\
$({\cal O}_{ll})_{ijkl} =$ & $(\overline{l_{{\rm R} i}} \gamma_\alpha l_{{\rm R} j})
(\overline{l_{{\rm R} k}} \gamma^\alpha l_{{\rm R} l})$ , 
\\
$({\cal O}_{Ll})_{ijkl} =$ & $(\overline{L_{{\rm L} i}} l_{{\rm R} j})(\overline{l_{{\rm R} k}} L_{{\rm L} l})$,  
\end{tabular}
\end{equation}
to three. These operators are also invariant under the exchange of the pairs of flavor indices 
$(ij) \leftrightarrow (kl)$ and under the permutation of the two flavor indices $i \leftrightarrow k$ 
and therefore, under the permutation of the other two flavor indices $j \leftrightarrow l$. 
What reduces the independent four-lepton operators ${\cal O}_{4l}$ to the set in Table 
\ref{mutaufourleptonoperators} if we also require flavor-parity conservation. 
Lepton number conservation follows from gauge invariance at this order.
Then, any NP coupling to muons and taus and heavy enough to evade direct observation 
at the LHC can be parametrized by a combination of the operators in this Table. 
We estimate the corresponding LHC and ILC reach assuming that only one of them 
has a non-vanishing coefficient at a time. 
The most stringent bounds are obtained for the four-muon operators 
$({\cal O}^{(1)}_{LL})_{\mu\mu\mu\mu}, ({\cal O}_{ll})_{\mu\mu\mu\mu}$ and 
$({\cal O}_{Ll})_{\mu\mu\mu\mu}$.
Using the same generation procedure as for ${\cal L}_{\rm eff}$ in
Eq. (\ref{Zprimeintegration})  
and applying the same cuts as in Table  \ref{table:Cutsoperators} (and
in Tables~\ref{3mu_cuts:table} and
\ref{4mu_cuts:table} except for the Mass window cut)
to the 3$\mu$ 
plus missing energy and 4$\mu$ samples, respectively, we obtain the
$95\%$ C.L. limits at the LHC~\footnote{The effective Lagrangian is
  defined as $x^i {\cal O}_i$,  with no additional global sign.}: 
\begin{eqnarray}
-10\mbox{ TeV}^{-2} &\leq x^{LL} &\leq 8.9\mbox{ TeV}^{-2}, \nonumber \\
-10.8\mbox{ TeV}^{-2} &\leq x^{ll} &\leq 10.4\mbox{ TeV}^{-2}, \\
-11.2\mbox{ TeV}^{-2} &\leq x^{Ll} &\leq 12.2\mbox{ TeV}^{-2},
\nonumber 
\end{eqnarray}  
and at the ILC:
\begin{eqnarray}
-38\mbox{ TeV}^{-2} &\leq x^{LL} &\leq 25\mbox{ TeV}^{-2}, \nonumber \\
-39\mbox{ TeV}^{-2} &\leq x^{ll} &\leq 27\mbox{ TeV}^{-2}, \\
-24\mbox{ TeV}^{-2} &\leq x^{Ll} &\leq 31\mbox{ TeV}^{-2}.  \nonumber
\end{eqnarray}  
Values which are similar to those derived for the heavy $Z'_{\mu -
  \tau}$ effective Lagrangian in Eq. (\ref{Zprimeintegration}), as  
similar are the applicable comments.  

If the leptophilic effective operators are allowed to involve
electrons, too, there are  
19 more independent four-lepton operators and an estimate of present 
bounds is reviewed in the Introduction.

\section{Conclusions}
\label{conclusions}

The outstanding performance of the LHC experiments makes them not only
discovery  
but also precision devices. With this in mind it is natural to
question whether the ILC is really a complementary machine,
especially if no signal of NP is established at the LHC. 
At any rate, the energy frontier is the domain of hadron machines, as
are the strong interactions,  
and at the end of the day even the EW gauge interactions, too. 
In general only the eventually huge backgrounds set the
limits of the LHC potential.  
Hence, an obvious question which we have addressed in this paper is if
NP only coupling to  
leptons and naturally accessible to the ILC is constrained at all by a
large hadron collider.   
The only lowest order interaction fulfilling these conditions is a new
leptophilic  
neutral vector boson. 

We have explored this SM addition allowing for general couplings to
muons and taus  
(and to their neutral counterparts). 
Couplings to electrons can be neglected because the corresponding LEP
limits make  
the LHC insensitive to them, as LFV bounds make the LHC insensitive to
leptophilic  
off-diagonal couplings. 
As a matter of fact, LHC experiments are only sensitive to the $Z'$
couplings to  
muons because couplings to taus can not be measured efficiently due to their 
larger backgrounds. 
The LHC reach for a leptophilic $Z'$ coupling to muons can be up to
$\sim$ 1 TeV 
for coupling constants of order one.
Moreover, the comparison of the 3$\mu$ plus missing energy and 4$\mu$
final states  
allows to determine the ratio of the RH singlet to the LH doublet muon couplings. 
In order to also determine the global normalization, the $Z'$ total width must be independently 
measured, which is not easy at the LHC (see Fig. \ref{LHCzprimewidth}). 
The corresponding determination of these couplings at the ILC relies on the 
4$\mu$, 2$\mu$ plus missing energy and 2$\mu$2$\tau$ samples 
(in the second and third sets two subsamples can be distinguished
depending on whether the
$Z'$ decays into $\mu^+ \mu^-$ or $\bar \nu \nu$ or into $\mu^+ \mu^-$ or $ \tau^+ \tau^-$, 
respectively), 
since the initial state is neutral and no $W$ can be exchanged in the
$s$-channel.  
The bounds which may be eventually derived are similar to the LHC ones, 
but only for light enough vector boson masses ($< 250$ GeV). 
In contrast, the $Z'$ couplings to taus can be only determined at the ILC 
in either case. 
However, the sensitivity of the ILC is very much dependent not only on how low 
the new vector boson mass is but also on the relative size of $g^{\prime \mu}_{\rm L, R}$ and 
$g^{\prime \tau}_{\rm L, R}$, because the former must be sizable for reconstructing 
the $Z'$ and the latter for producing a large enough signal to measure the tau couplings. 
This appreciably reduces the accessible parameter space at the ILC.   

In summary, the only departure from the SM eventually observable at the LHC, if a new 
leptophilic vector boson exists, is a moderate excess of events in the 3$\mu$ 
plus missing energy and 4$\mu$ samples with a $\mu^+ \mu^-$ pair peaking around the $Z'$ mass. 
The typical cross-section after preselection cuts being of few fb. 
On the other hand, the ILC can only confirm the observation of such a vector boson. 
Although for a rather light $Z'$ also coupling sizably to taus, an analysis of 
the 2$\mu$ plus missing energy and 2$\mu$2$\tau$ 
samples could also allow for the measurement of the $Z'$ couplings to taus. 
If this leptophilic vector boson would also couple to electrons, the leptonic processes 
$e^+ e^- \rightarrow \mu^+ \mu^- , \tau^+ \tau^-$ 
would offer the best place to look for such a new interaction \cite{delAguila:2011zs,Blas:2013ana}. 

We have also discussed the limit of a $Z'$ with a mass beyond the LHC
reach, that can be 
parametrized by the corresponding four-lepton operators. Limits on the independent 
invariant (dimension 6) operators involving four SM lepton multiplets 
can be estimated in a similar way, but the corresponding bounds on their 
coefficients, $\sim 10$ TeV$^{-2}$, are too weak 
to allow for a weakly coupled resonance interpretation. 

The fast simulation analyses have been performed using
\texttt{MADGRAPH 5} 
\cite{Alwall:2011uj}. 
The \texttt{UFO} model can be found in 
\href{http://cafpe.ugr.es/index.php/pages/other/software}
{http://cafpe.ugr.es/index.php/pages/other/software} 
in the package \href{http://cafpe.ugr.es/index.php/pages/other/software}
{Leptophilic_UFO.tar.gz}. 

\section*{Acknowledgements}

We thank J . Alcaraz, S. Bethke and T. Golling for useful discussions. 
MC would also like to thank the Institute for Theoretical Physics
at ETH Z\"urich for hospitality during the completion of this
project. 
This work has been supported in part by the European Commission 
through the contract PITN-GA-2012-316704 (HIGGSTOOLS), by the Ministry
of Economy and Competitiveness (MINECO), under grant numbers 
FPA2010-17915 and FPA2013-47836-C3-1/2-P, and by the Junta de Andaluc{\'\i}a 
grants FQM 101 and FQM 6552. 
M.C. was supported by the MINECO under the FPU program.

\providecommand{\href}[2]{#2}\begingroup\raggedright\endgroup
  
\end{document}